\documentclass[default,iicol]{sn-jnl}

\usepackage{graphicx}
\usepackage{changepage}
\usepackage{lipsum}
\usepackage{xcolor}
\usepackage{amssymb}
\usepackage{amsmath}
\usepackage{mathrsfs}
\usepackage[normalem]{ulem}  
\usepackage{url}
\usepackage{tabularx}
\usepackage{adjustbox}
\usepackage{booktabs}
\usepackage{array}
\usepackage{caption} 

\jyear{2021}%

\theoremstyle{thmstyleone}%
\theoremstyle{thmstyletwo}%

\theoremstyle{thmstylethree}%

\usepackage[ampersand]{easylist}
\newcolumntype{?}{!{\vrule width 1pt}}
\newcommand{\rt}[1]{{\textcolor{black}{{#1}}}}

\makeatletter
\def\thickhline{%
  \noalign{\ifnum0=`}\fi\hrule \@height \thickarrayrulewidth \futurelet
  \reserved@a\@xthickhline}
\def\@xthickhline{\ifx\reserved@a\thickhline
              \vskip\doublerulesep
              \vskip-\thickarrayrulewidth
             \fi
      \ifnum0=`{\fi}}
\makeatother

\newlength{\thickarrayrulewidth}
\begin{document}

\title[Figure/Caption Extraction]{The Digitization of Historical Astrophysical Literature with Highly-Localized Figures and Figure Captions}


\author*[1,2]{\fnm{Jill P.} \sur{Naiman}}\email{jnaiman@illinois.edu}

\author[3]{\fnm{Peter K. G.} \sur{Williams}}\email{pwilliams@cfa.harvard.edu}

\author[3]{\fnm{Alyssa} \sur{Goodman}}\email{agoodman@cfa.harvard.edu}

\affil*[1]{\orgdiv{School of Information Sciences}, \orgname{University of Illinois, Urbana-Champaign}, \orgaddress{\street{Daniel ST}, \city{Champaign}, \postcode{61820}, \state{Illinois}, \country{USA}}}

\affil[2]{\orgdiv{National Center for Supercomputing Applications}, \orgname{University of Illinois, Urbana-Champaign}, \orgaddress{\street{W Clark ST}, \city{Urbana}, \postcode{61801}, \state{Illinois}, \country{USA}}}

\affil[3]{\orgdiv{Center for Astrophysics}, \orgname{Harvard-Smithsonian}, \orgaddress{\street{Garden ST}, \city{Cambridge}, \postcode{02138}, \state{Massachusetts}, \country{USA}}}


\abstract{ Scientific articles published prior to the ``age of digitization" in the late 1990s contain figures which are ``trapped" within their scanned pages.  
    While progress to extract figures and their captions has been made, there is currently no robust method for this process. 
    We present a YOLO-based method for use on scanned pages, after they have been processed with Optical Character Recognition (OCR), which uses both grayscale and OCR-features.  
    We focus our efforts on translating the intersection-over-union (IOU) metric from the field of object detection to document layout analysis and quantify ``high localization" levels as an IOU of 0.9.
    When applied to the astrophysics literature holdings of the \rt{NASA} Astrophysics Data System (ADS), we find F1 scores of 90.9\% (92.2\%) for figures (figure captions) with the IOU cut-off of 0.9 which is a significant improvement over other state-of-the-art methods. }

\keywords{scholarly document processing, document layout analysis, astronomy.}


\maketitle

\section{Introduction}

With the rise of larger datasets and the ever increasing rate of scientific publication, scientists require the use of automated methods to parse these growing data products, including the academic literature itself.  
In addition to being a vital component of open science \cite{sandy2017,sohmen2018figures}, easily accessed and well curated data products are strongly encouraged as part of the submission process to most major scientific journals \cite{mayernik2017}.
However, data products in the form of figures, tables and formulas which are stored in the academic literature, especially from the ``pre-digital" era, published prior to $\sim$1997, are not accessible for curation unless methods are developed to extract this important information.

The extraction of different layout elements of articles is an important component of scientific data curation, with the accuracy of extraction of the elements such as tables, figures and their captions increasing significantly over the past several years \cite{icdar2017,podreview3,podreview1,lehenmeier2020layout}.
A large field of study within document layout analysis is the ``mining" of PDFs as newer PDFs are generally in ``vector" format -- the document is rendered from a set of instructions instead of pixel-by-pixel as in a raster format, and, in theory, the set of instructions can be parsed to determine the locations of figures, captions and tables \cite{klampfl2013unsupervised,bai2006automatic,choudhury2013figure}.

However, this parsing is non-trivial and many methods have been developed to complete this task. If the PDF's vector format is well structured, then text and images can be extracted by parsing this known PDF format.  Several packages exist which output text and/or images from such PDF files \cite{GROBID}.
Once such information is extracted, several methods for organizing of raw figures and text into figure-figure caption pairs, tables and other layout components (e.g. section headings, mathematical formulas) exist.
Historically, some of the most popular include heuristic methods where blocks of text are classified as figure or table captions based on keywords (like ``Fig." or ``Figure") \cite{Choudhury2013,pdffigures2}. 

Deep learning methods have become popular recently for vector and raster documents \cite{surveydeeplearning,deepfigures,sinha_rethinking_2022}, including those that use methods of semantic segmentation \cite{yang2017layout} and object detection \cite{saha2019}. 
When vector-PDFs are available\rt{,} these deep learning methods are often combined with heuristic methods to extract text during the layout analysis process \citep{deepfigures}.
While these methods are vital to the extraction of data products from recent academic literature, pre-digital literature is often included in digital platforms with older articles scanned at varying resolutions and deep learning methods developed with newer article training sets often perform poorly on this pre-digital literature \cite{scanbank}.
Additionally, layouts, fonts, and article styles are typically different for historical documents when compared to ``born-digital" scientific literature \cite{scanbank}.
In these cases, text extraction must be performed with optical character recognition (OCR), and figures and tables are extracted from the raw OCR results.
When applied to raster-PDF's with text generated from OCR, deep learning document layout analysis methods trained with newer or vector-based PDFs are often not as robust \cite{yang2017layout,scanbank}.
While progress has been made in augmenting these methods for OCR'd pages, especially for electronic theses and dissertations (ETDs) \cite{scanbank}, much work can still be done to extract layout elements from these older, raster-based documents. 

Large ``benchmark" raster-based datasets are available, however they tend to be comprised of a majority of newer articles.  For example, only about 2.6\% of the widely used PubLayNet dataset introduced in \cite{publaynet} are articles older than 1997 and benchmark datasets focused on historical scientific articles are less readily available \citep{scanbank}.
Additionally, definitions of what constitutes different layout elements -- figures, tables, and their captions -- can differ across datasets \citep{icdar2017,publaynet}, and large hand annotated benchmark datasets can suffer from inconsistencies in layout element definitions \citep{younas2019}.

In what follows, we outline a new methodology for extracting figures and figure captions from a dataset that includes both vector and raster based PDF's from the pre-digital 
scientific literature holdings of the NASA Astrophysics Data System (ADS)\footnote{\url{https://ui.adsabs.harvard.edu/}}.  
Our model applies deep learning object detection methods in combination with heuristic techniques to scans of article pages as well as the text features generated from processing scans through the \textsf{Tesseract} OCR engine \cite{tesseract} and combines the results from mining any vector based PDF's for their captions with \textsf{pdffigures2} \cite{pdffigures2} in a post-processing step.

While the focus of our model is the digitization of astronomical literature -- one of the original ``big data" sciences \cite{astrobigdata1,astrobigdata2} -- because our method relies heavily on features generated with OCR, our methodology is extendable to other \rt{large} scientific literature \rt{holdings which have already been OCR'd \citep[e.g.\ the HathiTrust U.S.\ Federal Document collection\footnote{\url{https://babel.hathitrust.org/cgi/mb?a=listis&c=2062901859}},][]{hathidata} }. 
Additionally, the design of our pipeline is heavily motivated by both the data (astronomical literature) and the expected users of our model (scientists and digital librarians).  Thus, we rely on open-source software (e.g. \textsf{Tesseract}) and programming languages used by both communities (e.g. \textsf{Python}). 
The outline of our paper is motivated by this possible wide range of utility: \autoref{section:design} details our dataset and outlines the design considerations of our pipeline,  \autoref{section:results} discusses our model architecture and accuracy and in particular \autoref{section:benchandgen} discusses the generalizability of our method to other ``benchmark" datasets.  
\rt{Especially relevant to other fields and the larger document layout analysis community is our discussion of the relationship of extraction metrics. In this discussion within \autoref{section:high}, we consider how the popular object detection metric intersection-over-union (IOU) can be used to quantify the information extracted from localized page objects.}
%
We outline our future plans for this work in \autoref{section:future}. All code is available on GitHub\footnote{Full project details for all work housed at: \url{https://github.com/ReadingTimeMachine}}.

\section{Design Considerations and Data Pre-processing} \label{section:design}


\subsection{The Data} \label{section:data}

\begin{figure}[!htp]
\centering
\includegraphics[width=0.45\textwidth]{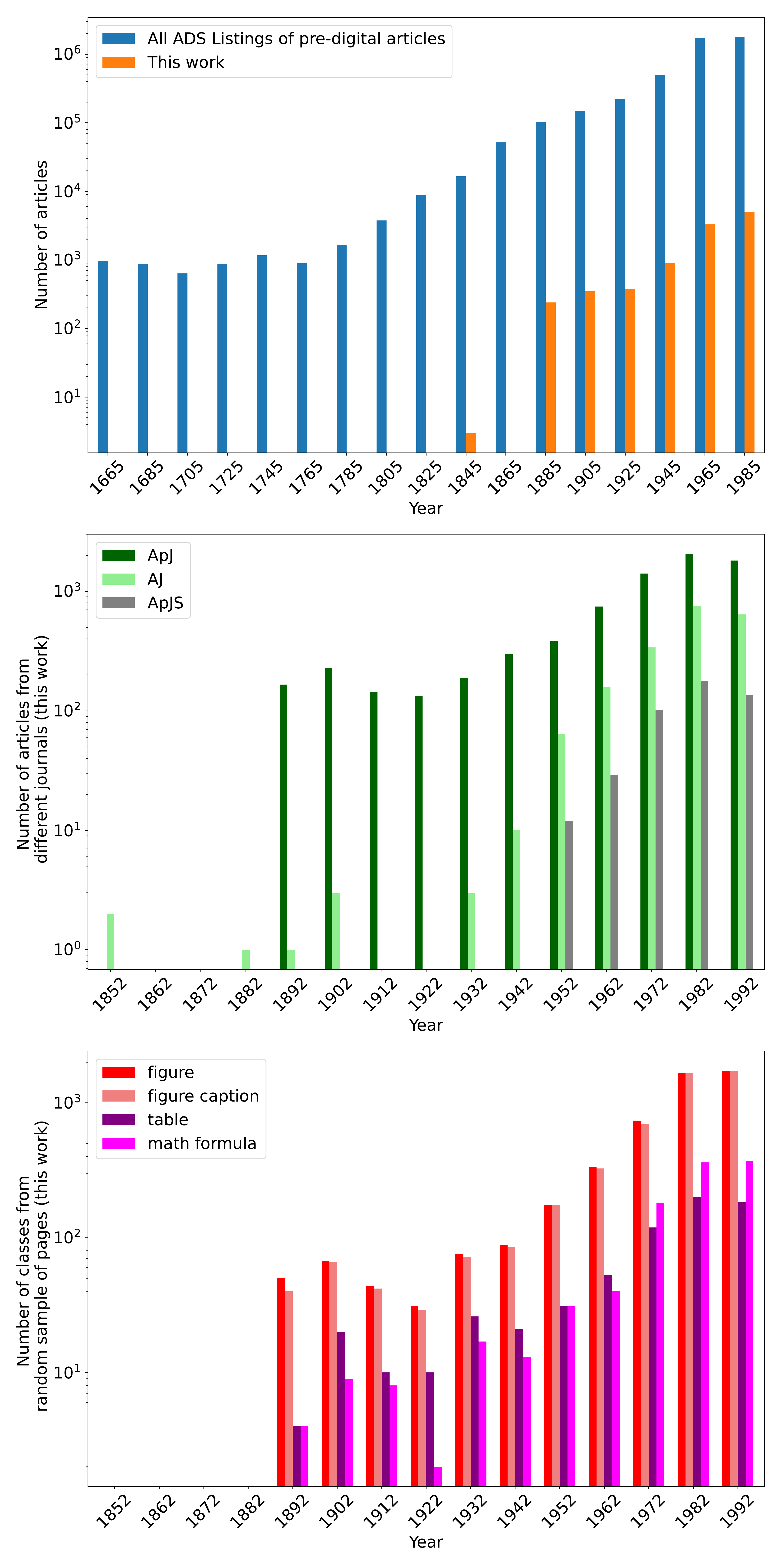}
\caption[The top panel shows the age distribution of all pre-digital ADS listings and the subset that constitutes our annotated database.  The middle panel shows the distribution of our full database across time and publisher for a total of $\sim$10k articles (out of the $\sim$56k of total pre-digital ADS listings from these three publishers).  The bottom panel shows the distribution of figures, figure captions, math formulas and tables per year in the main annotated subset of pages taken from articles in the full database. Totals across 
annotated subset of 5515 pages are: 5010 figures, 4925 figure captions, 1040 math formulas, 677 tables. There are 553 pages without additional page objects.]{The top panel shows the age distribution of all pre-digital ADS listings and the subset that constitutes our annotated database.  The middle panel shows the distribution of our full database across time and publisher for a total of $\sim$10k articles (out of the $\sim$56k of total pre-digital ADS listings from these three publishers).  The bottom panel shows the distribution of figures, figure captions, math formulas and tables per year in the main annotated subset of pages taken from articles in the full database. Totals across 
annotated subset of 5515 pages are: 5010 figures, 4925 figure captions\footnotemark, 1040 math formulas, 677 tables. There are 553 pages without additional page objects.}
\label{fig:distribution}
\end{figure}

The dataset used in this work is a subset of the English-language literature archived on the NASA Astrophysics Data System (ADS) 
of articles prior to the ``era of digitization" -- publishing year $\lesssim$ 1997 -- as shown in the top panel of \autoref{fig:distribution}.
Articles span the publications of The Astronomical Journal (AJ), The Astrophysical Journal (ApJ) including The Astrophysical Journal Supplement Series (ApJS) from the years of 1852-1997 (middle panel of Figure \ref{fig:distribution}).
Additionally, the dataset is also a subset of the literature featured in the Astrophysics Data System All-Sky Survey (ADSASS) which was an effort to associate each article with its place in the sky from the objects studied within the text \citep{adsass2012,adsass2015}. Chosen for this work are articles that are thought to contain images of the sky, as determined by heuristic determinations based on color distributions of pages \citep{adsass2012}.

\footnotetext{\rt{There are several full-page figures that do not have captions leading to a slightly higher number of figures than captions.}}

The bottom panel of \autoref{fig:distribution} shows the distribution of the subset of our larger database that is annotated with classes of figure, figure caption, table and math formula. 
These annotated pages were chosen randomly from all articles found with the heuristic methods of \citep{adsass2012}.
For this work, we will focus on the extraction of figures and figure captions, however we include the annotations for tables and math formulas in the downloadable data accompanying this paper as these elements are often of interest to the document layout analysis community \citep{iwatsuki2017,younas2019,publaynet}.  Our annotation process is discussed more fully in \autoref{section:annotations}.

The articles present in this work include those which are in a vector-PDF format (and therefore potentially parsable by \rt{``PDF-mining tools"}), and those that are in the raster-PDF format, with the majority of the articles in a raster-PDF format and the relative percentage of articles in vector-PDF format decreasing for older articles.  

\rt{Here, we define ``PDF-mining" as the process by which the set of ``instructions" which are used to construct vector-PDF documents is reverse-engineered to find the locations and content of page objects like text, tables and figures \citep{pdfact,pdfedit,pdffigures2,pdfminersix,pdftocairo,pypdf2,GROBID}.  In addition to these pure-heuristic methods, newer methods often make use of deep learning on rastered page images in combination with heuristics to locate and extract page objects which can often be more precise than pure heuristic methods alone \citep[e.g.][]{deepfigures}. However, due to their reliance on heuristics for text extraction, these newer methods are not accurate enough to extract page objects without access to the parsable instructions of vector-PDFs, and therefore have been shown to not be accurate on historical documents \citep{scanbank,scanbankthesis}.  In what follows, we limit our analysis to pure-heuristic based PDF mining software for simplicity, as the parsablity by these tools will give an estimation of how many of our articles are encoded with a set of instructions in the vector-PDF format.}

Determining whether or not an article is parsable by PDF mining tools requires a careful inspection of each PDF page, parsing outputs using mining software, and the quantification of how many words, figures, and tables are correctly parsed. 
For example, applying the PDF-image extraction tool \textsf{pdfimages}\footnote{\url{https://www.xpdfreader.com/pdfimages-man.html}} to several test pages generally results in corrupted output image files.  However, this process takes considerable time and computation to scale for all articles in our full dataset.  Thus, we must estimate parsability of the PDFs in other ways.

As a first \rt{\textit{estimation}} of parsability, we apply the PDF mining software \textsf{GROBID} \citep{GROBID} and \textsf{pdffigures2} \citep{pdffigures2} to the articles associated with our hand-annotated pages and calculate their ``parsability" with two metrics.  We look for parsed article outputs in which both figure and table numbers start at one and increase uniformly by one to their maximum figure and table number.  If these metrics do not hold it is likely because there is a missing or double-counted figure or table.  \rt{For these estimates, \textsf{GROBID} and \textsf{pdffigures2} are chosen as they are widely used PDF mining software \citep{lopez2009grobid,pdffigures2} and are used frequently to extract text and page objects specifically from scientific literature \citep{romary2015grobid,li2018extracting,yu2017convolutional}.}

\rt{In this analysis we assume the number system is either whole numbers (e.g.\ Figure 5 or Fig.\ 5) or roman numerals (e.g.\ Table IV or Tab.\ IV).  Objects which are tagged as tables or figures in \textsf{pdffigures2} which do not follow these numbering conventions account for 0.7\% of all objects.  In \textsf{GROBID} these non-standard numbering systems account for 1.3\% of all objects.  We further assume that these numbering systems are not mixed for a page object, but both may be operating in the same article, as tables are often enumerated with roman numerals while figures are counted with whole numbers. Thus, we define the PDF as parsable if \textit{either} the whole number or roman numeral system produces monotonically increasing integer figure and table numbers, each starting their count at one.}



\begin{figure}[!htp]
\centering
\includegraphics[width=0.45\textwidth]{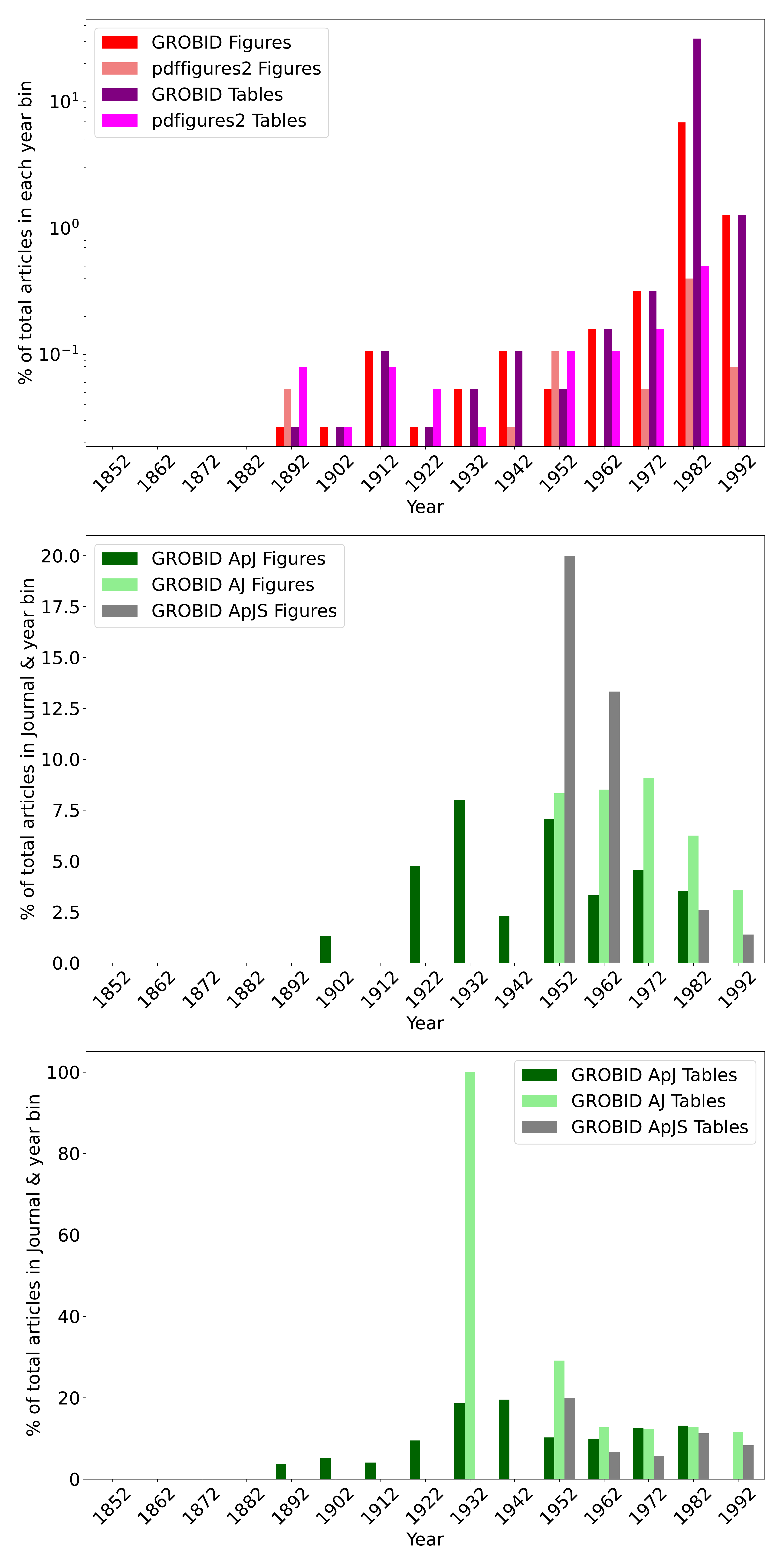}
\caption{PDF ``parsability" as estimated through the PDF-mining tools \textsf{GROBID} and \textsf{pdffigures2} \rt{in either whole numbers or roman numerals} as percentages of total articles in a 10-year bin.  An article is parsable if the sorted returned list of figures (tables) increase monotonically by one, starting at the figure (table) number ``1".  Estimates for parsability over all articles in our dataset are $\approx$\rt{1}-\rt{33}\% (see text for details).}
\label{fig:distributionPdfParse}
\end{figure}

The results of these estimates are shown in \autoref{fig:distributionPdfParse}.  \textsf{GROBID} and \textsf{pdffigures2} are able to parse articles best in the years $\approx$\rt{1945-1990} as shown in the upper panel of \autoref{fig:distributionPdfParse} with \textsf{GROBID} the more successful of the two in this time span.  In general, parsability as measured by tables is higher, which is to be expected as figures can often be labeled by words other than ``Fig" or ``Figure" (e.g. ``Plate" in this dataset) while tables are very rarely labeled with words other than ``Table" or ``Tab".  

The peak of \textsf{GROBID}'s parsing abilities from $\approx$\rt{1980}-1990 in figures (center panel of \autoref{fig:distributionPdfParse}) and tables (bottom panel of \autoref{fig:distributionPdfParse}) is likely due to the increase of articles from the Astronomical Journal (AJ) during this time (see center panel of \autoref{fig:distribution}) which appear to be \rt{comparable or} slightly more parsable than the other two journal formats.

Taken over all articles, the parsability \rt{using either whole numbers or roman numerals with}  \textsf{pdffigures2} for figures (tables) is $\approx$\rt{0.7}\% (\rt{1.1}\%).  For \textsf{GROBID} this percentage increases to $\approx$\rt{9.0}\% (\rt{33.7}\%) for figures (tables).  
\rt{Our estimation does not account for any erroneously discovered figures or tables.  The possibility of the addition of false positives to our parsability metrics will likely decrease the accuracies reported here for these tools.} \rt{Our accuracies are additionally likely lowered beyond the effects of lack of accounting for false positives as} our estimates do not include any checks for the correctness of the mined text. 

\rt{Finally, we remark here that there are many PDF parsers available beyond the three that we have tested here (\textsf{pdfimages}, \textsf{GROBID}, \textsf{pdffigures2}, with our focus on the latter two)  
which may provide further parsability improvements \citep{pypdf2,pdfact,pdfedit,pdfminersix,pdftocairo}. }

\subsection{Pipeline development}

Our final goal for this dataset is the hosting of figure-caption pairs on the Astronomy Image Explorer (AIE) database\footnote{\url{http://www.astroexplorer.org/}}.  Currently, a subset of the born-digital ADS holdings -- articles housed within the American Astronomical Society Journals (AAS) -- automatically populate AIE with their figure-caption pairs.
Thus, the pipeline described here begins with the initial OCR'ing of pages and ends with the extraction of figures and their captions by identifying the regions around the figures and the OCR'd words included in the caption region for hosting on a platform such as AIE.

\begin{figure}[!htp]
\centering
\includegraphics[width=0.5\textwidth]{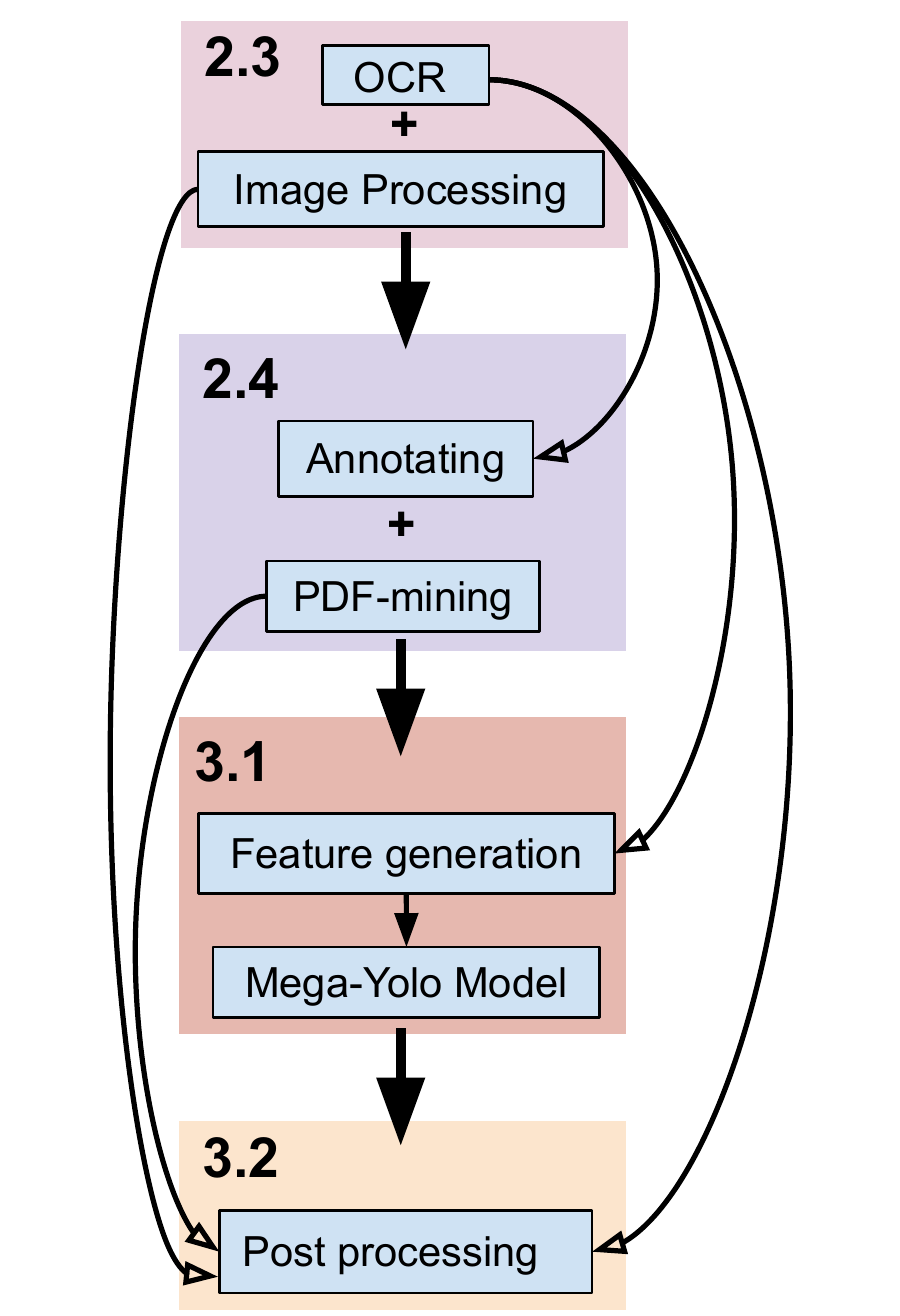}
\caption{Our overall pipeline is shown as four main steps. OCR and image processing is discussed in \autoref{section:imageprocessing}. Class definitions and ``codebook", \rt{along with} annotations and PDF mining \rt{are} discussed in \autoref{section:annotations}.  Feature selection and deep learning model description is housed in \autoref{section:modeldesign}, and post-processing techniques are discussed in  \autoref{section:postprocessing}. \rt{(See Appendix \ref{section:appendix} and \autoref{fig:pipelineexpand} for a larger breakdown of these steps.)} } 
\label{fig:pipeline}
\end{figure}

As the audience for this work is likely to be a mixture of scientists and digital librarians, we focus our efforts on developing a Python-based processing and training pipeline, as this is a language with great overlap between these two populations.
 \autoref{fig:pipeline} shows the outline of our full pipeline -- from data generation, through annotation, to training our deep learning model, and finally post processing the model results.

\subsection{OCR and Image Processing}\label{section:imageprocessing}

Our deep learning model makes use of OCR features for training (see \autoref{section:deeplearningmodel} for model details), which are generated by processing each page with \textsf{Tesseract}'s OCR engine \citep{tesseract}.
\rt{We use \textsf{Tesseract} parameters to find all English (lang=eng) text on the page, without any assumed format of lines or paragraphs (psm=12), the page rotation (OSD, psm=12) and to process the page with the LSTM OCR engine (oem=1).  These parameters allow \textsf{Tesseract} to find all text on the page independent of if the text is in a paragraph or on an image.  \textsf{Tesseract} does not locate or tag equations or images explicitly.  These parameters allow us to find the majority of text bounding boxes on a page accurately, however the resulting OCR text can be noisy and non-English characters are not included.  Cleaning of noisy, mixed-language, OCR text is a significant ongoing field of study in the document layout analysis community \citep{subramani_survey_2021,etter_synthetic_2019,boros_assessing_2022,ramirez-orta_post-ocr_2022} and is beyond the scope of this paper, but a subject of future work \citep{david,morgan}. }

Each randomly selected PDF page is processed into both a TIFF format (temporarily stored for OCR'ing) and JPEG format (for the extraction of gray-scale features later on in our pipeline).  
Original PDF articles are stored within ADS from high-resolution TIFF scans \citep{grant_nasa_2000}. As the PDF articles maintain the resolution of the original scans and are more fully supported for bulk download from ADS, we make use of these article formats for this work.
We extract both TIFF and JPEG from the article PDFs at high resolution (DPI=600) and resize the resulting images to a half their length and width to avoid pixelation and do not implement antialiasing.  TIFF images are used for the OCR process as there is evidence they produce fewer errors than other image formats \citep{choudhury2021}.
The temporary TIFF image is passed through \textsf{Tesseract} using a Python wrapper\footnote{\url{https://pypi.org/project/pytesseract/}} and utilizing Tesseract's optimization for OCR'ing full pages.  Outputs are stored in the \textsf{hOCR} format.
In this step basic image pre-processing is applied with \textsf{OpenCV} \citep{opencv} and its associated Python wrapper\footnote{\url{http://pypi.org/project/opencv-python/}} (thresholding and gaussian filtering) to address any artifacts on the page.

In conjunction \rt{with} the deep learning model, we use image processing techniques to heuristically find potential figure boxes as well.
The locations of the OCR'd words 
are used to mask out text and the modified pages are processed through a basic shape finder built with \textsf{OpenCV}, tuned to look for rectangles (four corners and sides comprised of approximately parallel lines).
This ``rectangle finder" is applied to several filtered versions of the page (histogram of oriented gradients, dilation, and color-reversal, and various levels of thresholding). The list of rectangles is culled with K-Means clustering on the locations of square corners, checking for artifact-rectangles which are small in size, and rectangles that are likely colorbars and not figures due to their aspect ratio.

OCR'ing a page and shape-finding with \textsf{OpenCV} takes approximately 20-25 seconds per page (tested on six cores of an Apple M1 Max with 64 Gb of RAM).



\subsection{Annotations and Class Definitions} \label{section:annotations}
Before delving into the details of our deep learning model, we consider several aspects of our annotation process that are necessary for a clear understanding of what our model is endeavoring to locate on each article page.

\begin{figure*}[!htp]
\centering
\includegraphics[width=1.0\textwidth]{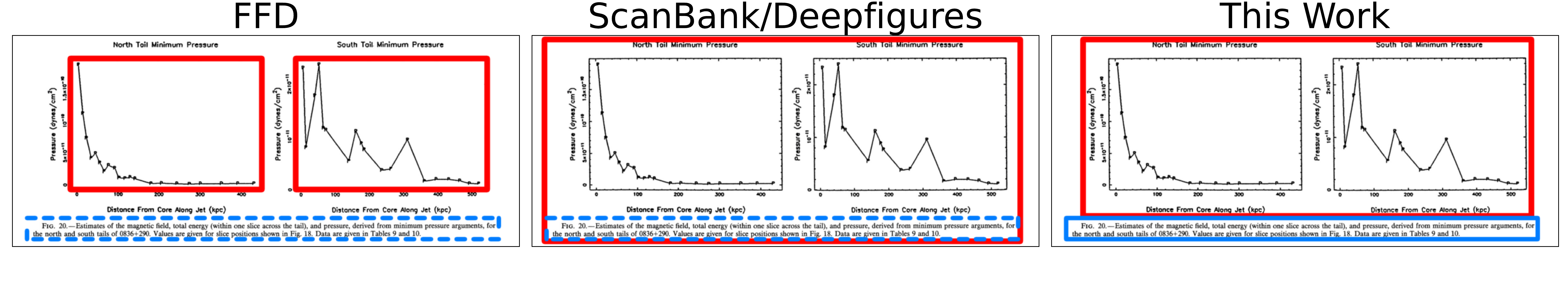}
\caption{Illustration of the ``codebook" for defining figures and figure captions (right) in comparison to other works (left, center). For example, some works \citep[e.g.][]{younas2019} (left \rt{red} boxes) typically split subfigures into separate figures while other works \citep{scanbank,scanbankthesis,deepfigures} (center \rt{red} box)
combine figure captions and figures into a single ``figure" box.  Caption boxes (shown in \rt{blue}) in other methods are delineated with dashed squares as applications often focus on detection of figures alone \citep[e.g.][]{scanbankthesis}.  Note as there is no overlap in datasets for direct comparison\rt{,} the boxes from other works are shown for illustrative purposes only. See \autoref{section:benchandgen} for a comparison of our category definitions and methodology with a subset of several of these datasets.} 
\label{fig:codebook}
\end{figure*}

We begin by defining the classes of figure and figure caption as there is often disagreement in the literature and occasionally between annotators of the same dataset \citep{icdar2017,younas2019}.
Here, as shown in \autoref{fig:codebook}, we define a figure as the collection of one or more panels on a single page which would be referred to as a single figure in a scientific article (i.e. ``Figure 3").  This is different than other works which often treat each ``sub-figure" as a separate figure \citep{younas2019}.
Additionally, figures are defined to include all axis labels and titles.
When figures are spread across multiple pages (often delineated with captions such as ``Fig 1b." or ``Figure 5, continued") each page of figures is classified as a separate figure.
If a figure caption extends horizontally further than its associated figure, the figure is extended horizontally to the edges of the figure caption (see magenta lines in \autoref{fig:postprocessing}).  These definitions retain the uniformity of other definitions (e.g.\ \cite{publaynet}) while defining figure and caption regions by non-overlapping boxes.
Finally, except in cases of unusual figure caption placement, the figure bounding boxes do not include the figure captions, in contrast to other definitions \citep[e.g.][]{icdar2017,publaynet,scanbank}, as part of our goal is to extract figure-caption pairs, we must delineate between these two different kinds of objects on the article page.

Figure captions are defined to be the caption text associated with these figure objects.  While there are many figures without captions, with the exception of a few cases, the majority of captions are on the same article page as their associated figures.  When more than one figure caption is potentially present (e.g. a sub-figure caption like ``Fig 4b") along with a longer figure caption, we choose the longer figure caption as the caption associated with the figure if it is on the same page as the figure.  If only a sub-figure caption is present on a page with a figure, we define this sub-figure caption as the caption of the figure.

Document layout objects are classified by hand using \textsf{MakeSense.ai} \citep{makesense} with the JPEG formatted images.
Once the hand annotation is completed\rt{,} checks using the OCR result are performed.  Specifically, the bounding boxes for the figure captions are modified to encompass the OCR word boxes as these are often offset or larger than the text presented in the original grayscale article page. 
When the OCR results are poor and do not capture the entirety of the caption, this will lead to an offset between visual and processed boxes.  
\rt{For example, if the skew of the page is significant, the scan too noisy, or the text too light, \textsf{Tesseract} may miss many words in the caption, leading to a bounding box that is significantly smaller than that which is detected visually by an annotator.  Even under ideal conditions, parts of individual letters can be truncated and therefore excluded from the OCR bounding box\footnote{\rt{This is a common issue with OCR engines and comes up often as a question online for \textsf{Tesseract} in particular, e.g.\ \url{https://stackoverflow.com/questions/57033120/bounding-boxes-around-characters-for-tesseract-4-0-0-beta-1}} \citep[e.g.][]{ocroffset}.}} 
However as we ultimately will be extracting the OCR'd text (for hosting on AIE in future work), we ignore these edge cases -- in our annotated dataset a single instance was reported. \rt{By including these noisy instances both here and in future annotation campaigns, our training data will include the noise that is expected to occur in OCR'd pages in the full article corpus. }
This is a crucial part of the annotation process to ensure we localize the text information of interest {\it and not a bounding box only on the grayscale image} which is often offset from the OCR results.

Finally, it is at this stage that we pass our PDF pages through the PDF mining software \textsf{pdffigures2} \citep{pdffigures2} to extract any figure and caption boxes for vector-PDFs.  
As we do not necessarily know \textit{a priori} which pages will be stored in vector-PDF\rt{,} we run \textsf{pdffigures2} on all pages.
Found figures and figure captions are stored for combination with our model's results in a post-processing step (see Step 3 in \autoref{section:postprocessing}).

Both modified hand-annotations and the results from PDF mining are stored in the YOLO-annotation style XML files \citep{bochkovskiy2020yolov4,Wang_2021_CVPR}.

\section{Deep Learning Model Design and Training} \label{section:deeplearningmodel}

In what follows, we discuss our development of a deep learning model that relies heavily on features derived from the OCR results of raster-PDF's to make use of the preponderance of these types of PDFs in our dataset.  



\subsection{Model Design and Feature Selection} \label{section:modeldesign}

Typical modern methods rely on deep learning techniques to detect layout elements on pages \cite{surveydeeplearning}, often in combination with heuristics \cite{deepfigures}.
Methods span the range of object detection using models like YOLO \cite{yolo1,deepfigures,scanbank} to, more recently, Faster R-CNN \cite{fasterrcnn,fasterrcnnDeepDesert,vo2018,younas2019,fasterrcnntables} and Mask-CNN \cite{maskrcnn,maskrcnnDocbank,maskrcnncdec}. Additionally, several pixel-by-pixel segmentation models have been proposed using semantic segmentation \cite{yang2017layout} and fully convolutional networks \cite{fcnn,fcnncharts}, including ``fully convolutional instance segmentation" \cite{fcis1,fcis2,fcis3}.

Often these models employ a variety of features derived from article pages as inputs along side or in place of the unprocessed page.
Some of the more popular recent methods leverage image processing and computer vision techniques \cite{younas2019} including connected component analysis \cite{connectedComp,conncomp2,conncomp3,conncomp4} and the distance transform \cite{fasterrcnntables} or some combinations thereof \cite{fifo}. 

While the aim of many methods is to detect page objects before any OCR process \rt{\cite[e.g.][]{podreview1,anotherDLAreview2022,binmakhashen_document_2019,kosaraju_document_2019}}, here we implement methods that can be applied after \rt{in an effort to support future extensions of our work to large digital archives which are constructed from previously-OCR'd articles such as those within the HathiTrust and the historical documents in the Internet Archive \citep{christenson2011hathitrust,ribaric2009automatic,iacor}.}
In what follows, we use a \textsf{Tensorflow} implementation of YOLO-v5\footnote{\url{https://github.com/jahongir7174/YOLOv5-tf} \\ \url{https://github.com/jmpap/YOLOV2-Tensorflow-2.0}} \cite{yolov1,yolov5} and focus our efforts on feature exploration by utilizing a set of features derived from the OCR outputs themselves with the goal to choose the smallest number of the ``best" features on which to build our model.  

In addition to the raw grayscale page, there are several possible features derived primarily from the \textsf{hOCR} outputs of \textsf{Tesseract}.
To minimize storage, each feature is scaled as an unsigned-integer, 8-bit ``color" channel in a 512x512 (pixels), multi-layer image which is fed into a ``mega-YOLO" model capable of processing more than three color channels.  Features explored which are output from \textsf{Tesseract} in  \textsf{hOCR} format  include:
\begin{itemize}
    \item {\it{fontsize (fs)}}: the fontsize for each word bounding box is normalized by subtraction of the median page fontsize and division by the standard deviation.  Bounding boxes with fonts outside five standard deviations are ignored.
    \item {\it carea ($c_{\rm b}$)}: the ``content area" from automatic page segmentation includes large blocks of text and sometimes encapsulates figures into separate ``content areas", but not consistently.
    \item {\it paragraphs ($p_{\rm b}$)}: automatically segmented groups of words as likely paragraphs. Often overlaps with ``carea".
    \item {\it ascenders (asc)}: from typography definitions -- the amount of letters in the word that are above the letter ``caps" (e.g. the top of the letter ``h").  Ascenders are normalized by subtracting the median value for each page.
    \item {\it descenders (dec)}: a typographical element -- the amount of letters in the word that are below the letter ``bottoms" (e.g. the bottom curl of the letter ``g").  Descenders are normalized by subtracting the median value for each page.
    \item {\it word confidences (wc)}: the percent confidence of each word 
    \item {\it word rotation (t$_{\rm ang}$)}: rotation of word in steps of 0$^\circ$, 180$^\circ$, and 270$^\circ$.
\end{itemize}

\noindent Other features derived from the page scan and OCR-generated text are:
\begin{itemize}
    \item {\it grayscale (gs)}: the image is collapsed into grayscale using the page's luminance. The majority of images are already in grayscale and those few that are in color are transformed to grayscale.
    \item {\it fraction of letters in a word (\%l)}: the percentage of characters in a word that are letters (scaled 125-255 in order to preserve a ``true zero" for spaces in the scanned page that contain no words).
    \item {\it fraction of numbers in a word (\%n)}: the percentage of \rt{characters} in a word that are \rt{numbers} (scaled 125-255).
    \item {\it punctuation (p)}: punctuation marks are tagged as 250, non-punctuation characters are tagged as 125 (saving 0 for empty, non-word space). 
    \item {\it spaCy POS (SP)}: spaCy's \cite{spacy2} 19 entries for ``part of speech" (noun, verb, etc.) in the English language
    \item {\it spaCy TAG (ST)}: more detailed part of speech tag with 57 values
    \item {\it spaCy DEP (SD)}: the 51 ``syntactic dependency" tags which specify how different words are related to each other
\end{itemize}

\begin{figure*}[!htp]
\centering
\includegraphics[width=1.0\textwidth]{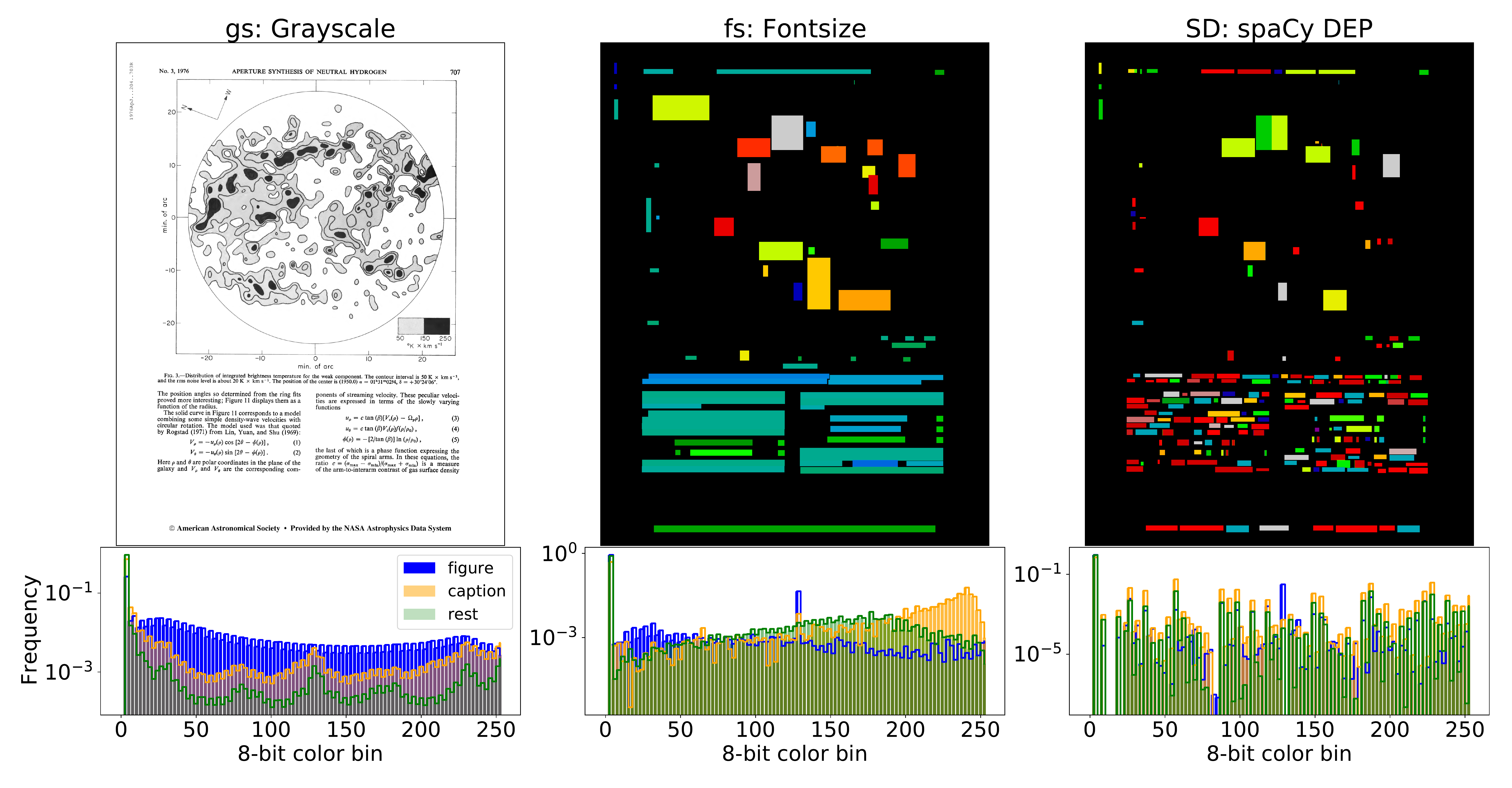}
\caption{Examples of selected features for a single page (top row) and the distribution across all pages in the annotated dataset (bottom row).  All features have been rescaled into 8-bit color bins (see \autoref{section:modeldesign}). Here, ``rest" refers to rest of the page that does not include figures or captions.  Several features show clear differences in distributions (e.g. grayscale and fontsize) while others do not (e.g. spaCy DEP).  \rt{Note: for plotting purposes we have left the grayscale image un-inverted, but the image is inverted when used as a feature in our model.}}
\label{fig:features}
\end{figure*}

\autoref{fig:features} shows an example of a selection of these features (grayscale ({\it gs}), fontsize ({\it fs}), and spaCy DEP ({\it SD})) for a single page and their distributions across all pages in our annotated dataset.  \rt{For illustrative purposes, we have left the grayscale image in the upper left panel of \autoref{fig:features} un-inverted, however for use as a feature the grayscale is inverted, as is typical for document layout applications \citep[e.g.][]{maskrcnnDocbank}.  The top panels of \autoref{fig:features} showcase the different regions highlighted by different feature layers.  While the grayscale (top left panel) allows a reader to localize the figure visually, the region of the figure is also denoted by the large fontsizes of the OCR words that are picked out within the figure (red and yellow blocks, top middle panel).  While often times erroneous detections, these large-fontsize words are typical for those found by \textsf{Tesseract} within figures.  Additionally, the caption located directly below the figure is highlighted by smaller (bluer) fontsizes than the rest of the text on the page (majority green).  These relatively obvious separate regions in fontsize are not replicated in the patterns seen in the spaCy DEP parameter (top right panel). Here, the colors of OCR bounding boxes do not follow any obvious pattern between regions of the figure, its caption, or the other text on the page.}

\rt{These patterns extend to the distributions of these three features across all scanned pages in our dataset, as depicted by the histograms in the lower panels of \autoref{fig:features}.  }
The example features of grayscale and fontsize show differences in distributions in the three categories of figure, figure caption and the ``rest" of the page -- grayscale distributions are more uniform inside figures (\rt{bottom} left \rt{histogram} of \autoref{fig:features}) and figure captions show a peak in the fontsize distributions toward higher values when compared to the fontsize distributions of figures (\rt{bottom} middle \rt{histogram} of \autoref{fig:features}). Trends in other features are harder to determine, as illustrated in the bottom right \rt{histogram} of \autoref{fig:features} which shows a less clear distinction between figures, figure captions, and the rest of the page for the feature of spaCy DEP.  

\rt{In addition to potentially being able to distinguish types of page regions, these features are similar in morphology to computer vision features such as connected components \cite{connectedComp,conncomp2,conncomp3,conncomp4}, making them a natural extension of such work (see for example Figure 2 of \cite{younas2019} to compare to our \autoref{fig:features}).}

In \autoref{section:results} we discuss our best model which includes (grayscale,  ascenders, descenders, word confidences, fraction of numbers in a word, fraction of letters in a word, punctuation, word rotation and spaCy POS) as the set of input features.




\subsection{Post-Processing Pipeline} \label{section:postprocessing}

After features are selected and the model is trained we modify the final found boxes by merging them with OCR word and paragraph boxes and any heuristically found captions and figures at the fractional-pixel level (results are rounded to nearest pixel for intersection-over-union (IOU) calculations to match precision of ground truth boxes).

Post-processing is a common practice in document layout analysis \cite{fifo,yi2017}, however it often differs between implementations and is occasionally not incorporated into a final pipeline \cite{wu2019detectron2}.
\begin{figure*}[!htp]
\centering
\includegraphics[width=0.95\textwidth]{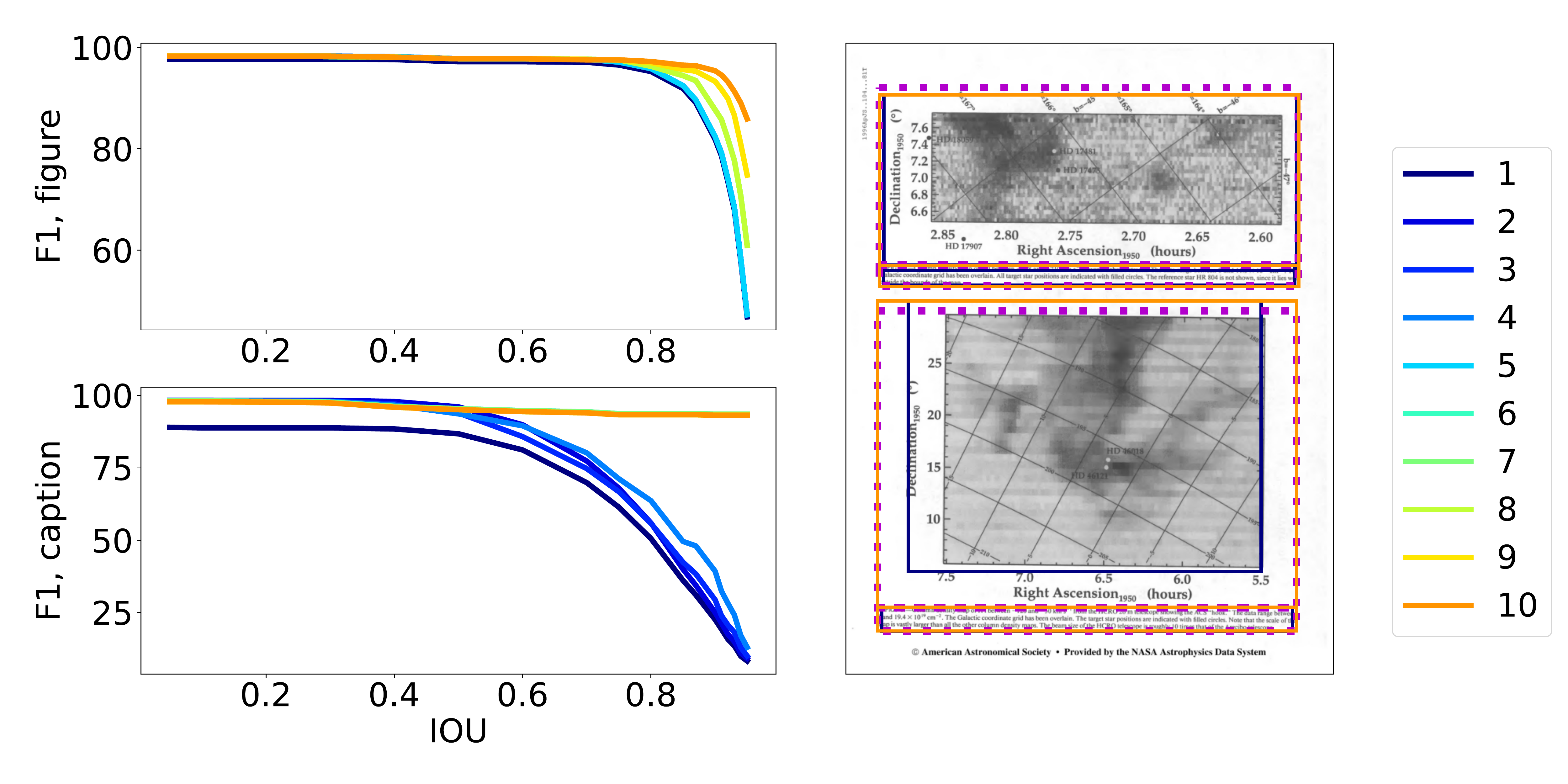}
\caption{Effects of post-processing steps on F1 (left plots) for Model 12 (m12 in \autoref{tab:features} and \autoref{tab:featuresSelection}).  Post-processing drives changes in the metrics at larger IOU's -- IOU$\gtrsim$0.8 and IOU$\gtrsim$0.6 for figures and captions, respectively.  Changes are depicted for a single page (right plot) showing initial found boxes (Step 1, dark blue) and final (Step 10, orange) in comparison to true boxes (thick \rt{dashed} magenta).}
\label{fig:postprocessing}
\end{figure*}
\autoref{fig:postprocessing} depicts how found boxes and 
F1 score changes with each post-processing step in our pipeline when we compare ground-truth (true) boxes to model-found boxes at various post-processing steps:
\begin{itemize}
    \item Step 1: ``raw" found boxes are those culled with non-maximum suppression \rt{to combine several overlapping proposed boxes (i.e.\ regions that are tagged with a high probability of containing a page object of a particular class) into a single bounding box \citep{yolo1,nms1,nms2,nms3,nms4,nms5}. } 
    \item Step 2: if two found boxes overlap with an IOU~$\ge$~0.25 the box with the lowest score is removed, decreasing false positives (FP)
    \item Step 3: \textsf{pdffigures2}-found figure caption boxes replace those found with the deep learning model when they overlap\rt{,} which increases caption true positive (TP) rate and decreases FP and false negative (FN) at large IOU thresholds.
    \item Step 4: \rt{c}aption boxes are found heuristically by first applying a gaussian blur filter on an image of filled-in OCR text boxes. Contours of this image that overlap with text boxes that match with a fuzzy-search of words such as ``Fig.", ``Figure" and ``Plate" are labeled as heuristically-found.  If a heuristically-found caption box overlaps with a mega-YOLO-found box, we take the top of the heuristic box (which tends to be more accurate) and the minimum (maximum) of the left (right, bottom) of the two boxes.  This results in an overall increase in TP while FN and FP drop.  
    \item Step 5: found captions are expanded by their overlap with OCR word and paragraph boxes, allowing for multiple ``grow" iterations in the horizontal direction.  Found boxes are only expanded by paragraph and word boxes if the centers of paragraph and word boxes overlap with the found box.
    \item Step 6: if found figure boxes overlap with rectangles that are found through image processing (as described in \autoref{section:imageprocessing}), the found box is expanded to include the image processing rectangle.  This increases the TP rate at larger IOU thresholds for figures.
    \item Step 7: any found captions that have areas larger than 75\% of the page area are discarded leading to a slight drop in FP for captions.
    \item Step 8: \rt{c}aptions are paired to figures by minimizing the distance between caption center and bottom of a specific figure.  Rotational information from the page and overall rotation of the OCR words is used to determine the ``bottom" of the figure. Any captions without an associated figure on a page are dropped, leading to a drop in FP.
    \item Step 9: found figure boxes are extended down to the tops of their associated captions increasing TP for figures and captions at high IOU thresholds.
    \item Step 10: if a figure caption extends horizontally further than its associated figure, the figure is extended horizontally to the edges of the figure caption.  This leads to an increase in TP rates for figures at high IOU thresholds.
    
    \end{itemize}

Steps 9 and 10 are similar to the steps described for annotated boxes in \autoref{section:annotations}. The effects \rt{on} the metrics shown in Figure \ref{fig:postprocessing} are modest and predominately affect the results at high IOU thresholds (IOU$\gtrsim$0.9) for figures.

\subsection{Feature Selection Ablation Experiments } \label{section:experiments}
To determine the set of features which produce the most accurate model while minimizing the feature memory footprint, we conduct a series of ablation experiments, summarized in \autoref{tab:features} and \autoref{tab:featuresSelection}.
In all feature selection runs we use 75\% of our data in the training set, 15\% in validation, and 10\% in the test dataset.  Results in \autoref{tab:featuresSelection} are shown for this feature selection test dataset.

\begin{table*}[!htp]
\small
\begin{center}
\begin{tabular}{|c|l|}
    \hline
    model & Description \\
    \thickhline 
    m1 & gs  \\
    \hline
    m2 & gs + fs \\
    \hline
    m3 & gs + fs + asc + dec \\
    \hline
    m4 & gs + fs + asc + dec + wc \\
    \hline
    m5 & gs + fs + asc + dec + wc + \%n  + \%l + p\\
    \hline
    m6 & gs + fs + asc + dec + wc + \%n  + \%l + p + $\rm{t_{ang}}$  \\
    \hline
    m7 & gs + fs + asc + dec + wc + \%n  + \%l + p + $\rm{t_{ang}}$ + SP  \\
    \hline
    m8 & gs + fs + asc + dec + wc + \%n  + \%l + p + $\rm{t_{ang}}$ + SP + ST + SD   \\
    \hline
    m9 & gs + fs + asc + dec + wc + \%n  + \%l + p + $\rm{t_{ang}}$ + SP + ST + SD + $\rm{p_{b}}$  \\
    \hline
    m10 & gs + fs + asc + dec + wc + \%n  + \%l + p + $\rm{t_{ang}}$ + SP + ST  + SD + $\rm{p_{b}}$ + $\rm{c_{b}}$  \\
    \hline
    \thickhline
    m11 & gs + fs + wc +  \%n  + \%l  + p + $\rm{t_{ang}}$ + SP  \\
    \hline
    \bf m12 & \bf gs + asc + dec + wc + \%n  + \%l  + p + $\rm{t_{ang}}$ + SP  \\    
    \hline
    m13 & gs + asc + dec + wc + \%n  + p + $\rm{t_{ang}}$ + SP  \\    
    \hline
    m14 & gs + asc + dec + wc + \%n  + \%l  + $\rm{t_{ang}}$ + SP  \\    
    \hline
    m15 & gs + asc + dec + wc + \%n  + \%l  + p + $\rm{t_{ang}}$ \\    

    \hline
\end{tabular}
\end{center}
\caption{ Ablation experiments with the features discussed in \autoref{section:deeplearningmodel}. All models include post-processing (\autoref{section:postprocessing}).  Our ``best" model, as determined by metrics in \autoref{tab:featuresSelection} and the discussion of \autoref{section:experiments}, is Model 12 (m12) highlighted in bold.} 
\label{tab:features}
\end{table*} 

As it is computationally prohibitive to test all combination of all fourteen different features, we first adopt the strategy of including sets of one or two groups of features at a time until we have a model containing all fourteen features, as shown above the thick horizontal line in \autoref{tab:features}.

\rt{Similar to other work \citep[e.g.][]{yang2017layout}, these parameter combinations endeavor to follow an ``intuitive" build up of complexity for our model. For example, our first model (m1) which consists of using only the grayscale (gs) image as a feature, mimics simply applying YOLO to an unprocessed page. We then add ``primary" OCR features, which we define here as those which come directly from the OCR engine without extra processing.  As the fontsize (fs) of objects like figure captions and figure axis labels is typically visually different from the main article text (see \autoref{fig:features}), we add this feature second in m2, followed by ascenders (asc) and descenders (dec) which can also change with the different fonts which are often present in different text objects (m3).  Word confidences (wc) can be lower for the spurious OCR detections that can often occur within figures which we add to our next model (m4).}

\rt{Secondary OCR features are defined here as features which are derived from the words generated by the OCR engine.  The proportion of OCR words which are numbers and letters and characters which are punctuation (\%n, \%l and p, respectively) typically changes between caption or main-text words and OCR words found on axis labels or inside figures, as does the rotation of words (m5 and m6). Even further removed from the raw OCR data are the linguistic features derived from the found OCR words (spaCy SP, ST, SD in m7-m9), and finally groupings of words into paragraphs and large word blocks (p$_{\rm b}$ and c$_{\rm b}$ in m10).}

From these ten models, we select the most accurate, defined here as having a high F1 score for both figures and their captions, while maintaining a low false positive score (FP).  Model 7 is the ``best" model out of these first ten models in \autoref{tab:featuresSelection}. 
We then subtract one or two features from this model in combinations shown below the thick horizontal line in \autoref{tab:features}.  Using the same selection criteria leads us to choose Model 12 as our overall ``best" model which includes the features of (grayscale,  ascenders, descenders, word confidences, fraction of numbers in a word, fraction of letters in a word, punctuation, word rotation and spaCy POS) as highlighted in \autoref{tab:featuresSelection}. \rt{This model represents a combination of not only the grayscale page, but many of the primary OCR features which intuitively would differ between regions of main-text, figures and figure captions.  The addition of the secondary feature of spaCy's part of speech tag (SP) suggests a likely difference in the word usage between regions of text which align with previous work \citep{yang2017layout,transformersementic,sciwordembedd}.}

The implemented optimizer is Adam with a $\beta_1=0.937$ $\beta_2=0.999$. 
Learning rate is scheduled using a cosine scheduler which depends on initial learning rate, number of epochs and batch size. Practically, when applied to our model this results in a linear increase in learning rate by a factor of $\sim$1.6 in the initial epoch (flat after). Our optimal initial learning rate of 0.004 was chosen from a small set of learning rates (0.008, 0.004, 0.0004, 0.0002). All experiments are run for 150 epochs and converge within this time (tracked by validation losses). No data augmentation is applied. Training is performed on a Tesla V100-SXM2 GPU with an average time of $\sim$6.5~minutes per epoch.

\begin{table*}[!htp]
\footnotesize
\begin{center}
\begin{tabular}{|c?cc|cc|cc?cc|cc|cc|}
    \hline
    & \multicolumn{2}{c|}{TP}  & \multicolumn{2}{c|}{FP} & \multicolumn{2}{c?}{FN} & \multicolumn{2}{c|}{Prec} & \multicolumn{2}{c|}{Rec} & \multicolumn{2}{c|}{F1} \\
     & fig & cap & fig & cap & fig & cap & fig & cap & fig & cap & fig & cap \\

    \hline

    m1 & 90.3\unskip & 88.3\unskip &  11.3\unskip & 10.0\unskip & 1.8\unskip & 4.3\unskip & 88.9\unskip & 89.8\unskip & 98.0\unskip & 95.4\unskip & 93.3\unskip & 92.5\unskip \\

    m2 & 89.5\unskip & 86.9\unskip &  10.9\unskip & 8.6\unskip & 2.6\unskip & 6.4\unskip & 89.2\unskip & 91.0\unskip & 97.2\unskip & 93.2\unskip & 93.0\unskip & 92.1\unskip \\

    m3 & 85.9\unskip & 86.9\unskip &  17.9\unskip & 12.5\unskip & 1.8\unskip & 5.1\unskip & 82.8\unskip & 87.4\unskip & 97.9\unskip & 94.4\unskip & 89.7\unskip & 90.8\unskip \\

    m4 & 90.5\unskip & 88.7\unskip &  10.1\unskip & 8.8\unskip & 2.0\unskip & 3.7\unskip & 90.0\unskip & 91.0\unskip & 97.8\unskip & 96.0\unskip & 93.7\unskip & 93.4\unskip \\

    m5 & 84.5\unskip & 87.3\unskip &  15.7\unskip & 7.2\unskip & 2.2\unskip & 7.0\unskip & 84.3\unskip & 92.4\unskip & 97.4\unskip & 92.6\unskip & 90.4\unskip & 92.5\unskip \\

    m6 & 89.5\unskip & 89.1\unskip &  11.9\unskip & 9.0\unskip & 2.0\unskip & 4.3\unskip & 88.3\unskip & 90.8\unskip & 97.8\unskip & 95.4\unskip & 92.8\unskip & 93.0\unskip \\

    m7 & 92.8\unskip & 88.1\unskip &  8.0\unskip & 9.0\unskip & 1.4\unskip & 4.1\unskip & 92.0\unskip & 90.7\unskip & 98.5\unskip & 95.6\unskip & 95.1\unskip & 93.1\unskip \\

    m8 & 90.5\unskip & 90.0\unskip &  9.1\unskip & 7.2\unskip & 2.0\unskip & 3.9\unskip & 90.9\unskip & 92.6\unskip & 97.8\unskip & 95.9\unskip & 94.2\unskip & 94.2\unskip \\
    
    m9 & 84.3\unskip & 84.2\unskip &  13.5\unskip & 7.4\unskip & 4.0\unskip & 9.4\unskip & 86.2\unskip & 91.9\unskip & 95.4\unskip & 89.9\unskip & 90.6\unskip & 90.9\unskip \\
    
    m10 & 88.7\unskip & 87.5\unskip &  11.5\unskip & 9.6\unskip & 1.8\unskip & 4.3\unskip & 88.6\unskip & 90.1\unskip & 98.0\unskip & 95.3\unskip & 93.0\unskip & 92.6\unskip \\
    m11 & 90.5\unskip & 92.4\unskip &  10.5\unskip & 6.8\unskip & 0.8\unskip & 1.8\unskip & 89.6\unskip & 93.2\unskip & 99.1\unskip & 98.0\unskip & 94.1\unskip & 95.6\unskip \\
    \bf m12 & \bf 92.2\unskip & \bf 89.1\unskip &  \bf 6.4\unskip & \bf 6.6\unskip & \bf 2.4\unskip & \bf 4.9\unskip & \bf 93.5\unskip & \bf 93.1\unskip & \bf 97.4\unskip & \bf 94.8\unskip & \bf 95.4\unskip & \bf 94.0\unskip \\
    m13 & 92.8\unskip & 88.7\unskip &  7.8\unskip & 8.4\unskip & 2.0\unskip & 4.3\unskip & 92.2\unskip & 91.4\unskip & 97.9\unskip & 95.4\unskip & 95.0\unskip & 93.3\unskip \\

    m14 & 87.3\unskip & 88.7\unskip &  15.9\unskip & 8.4\unskip & 1.2\unskip & 6.4\unskip & 84.6\unskip & 91.4\unskip & 98.6\unskip & 93.3\unskip & 91.1\unskip & 92.3\unskip \\

    m15 & 89.9\unskip & 89.5\unskip &  8.7\unskip & 5.9\unskip & 2.4\unskip & 5.1\unskip & 91.2\unskip & 93.8\unskip & 97.4\unskip & 94.6\unskip & 94.2\unskip & 94.2\unskip \\

    \hline
\end{tabular}
\end{center}
\caption{Metrics for models described in \autoref{tab:features}. There are {\protect497\unskip} figures (fig) and {\protect488\unskip} figure captions (cap) used in the feature selection test dataset. IOU is 0.9 for both figures and captions.   
TP, FP and FN are shown as percentages of the total instances.
}
\label{tab:featuresSelection}
\end{table*}


\section{Results} \label{section:results}

\begin{figure}[!htp]
\centering
\includegraphics[width=0.45\textwidth]{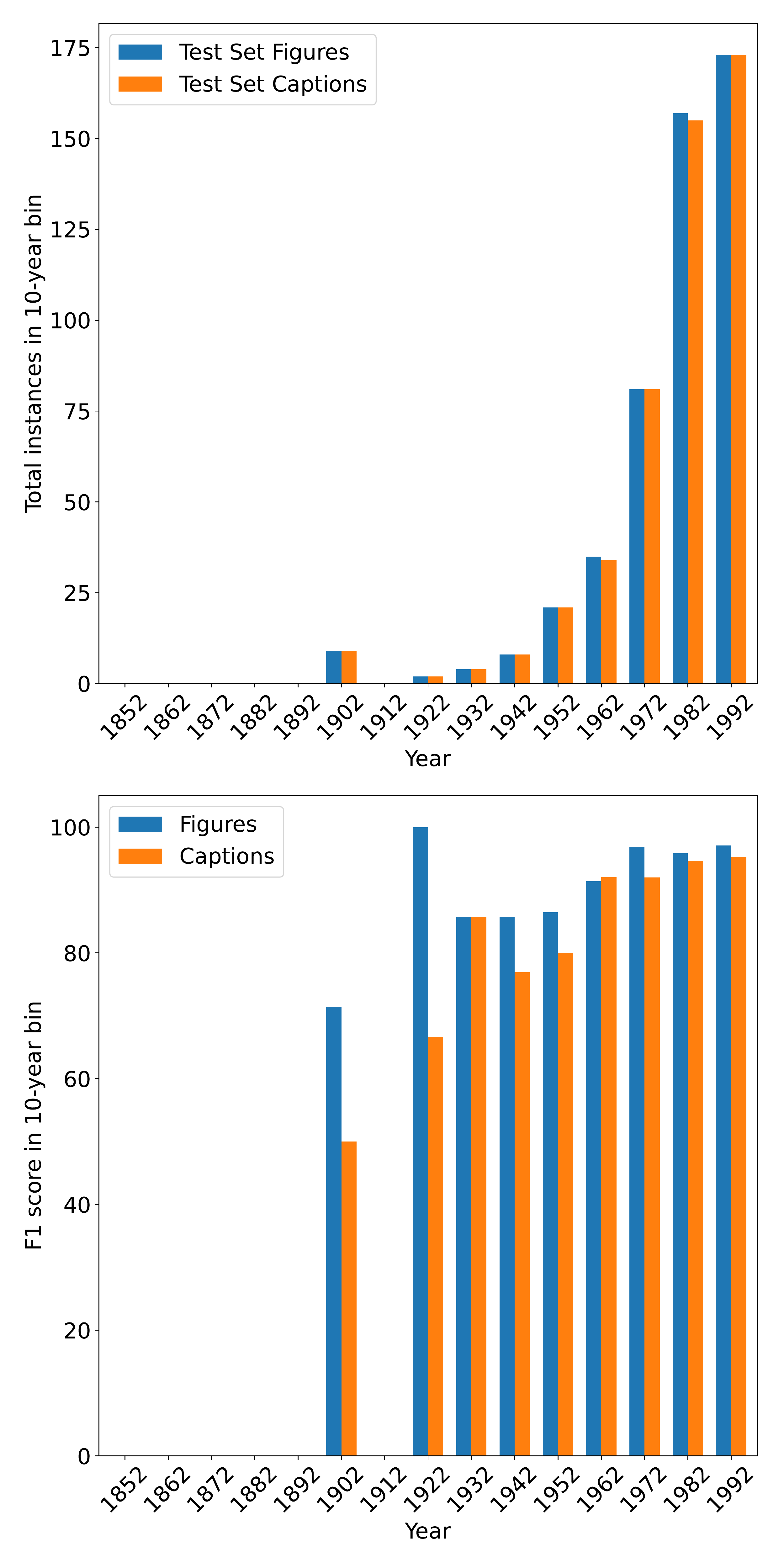}
\caption{Distribution of figures and figure captions in the final test dataset (\textit{top panel}) and the distribution of F1 scores in each time bin (\textit{bottom panel}) for an IOU=0.8 threshold.  Bins are chosen to match the bottom two panels of \autoref{fig:distribution} and all panels in \autoref{fig:distributionPdfParse}. However, there are no pages from articles prior to 1903.}
\label{fig:f1time}
\end{figure}

To quantify the results of our ``best" model (Model 12) on un-seen data we annotate an additional $\approx$600 pages as a ``final test dataset" of $\approx$500 figure and figure caption ground-truths (490 and 487, respectively).  Including post-processing, evaluation takes on average 1.8~seconds per page on a single core of an Apple M1 Max with 64 Gb of RAM.
The distribution of figures and figure captions in 10-year time bins for this dataset \rt{is} shown in the top panel of \autoref{fig:f1time}.

Metrics for the performance of our model in this final test dataset across several IOU cut-offs are shown in \autoref{tab:mainResults}.
While true positives (TP) and false negatives (FN) are relatively flat across all IOU cut-offs for figures, false positives increase by a small factor at the IOU=0.8 cut-off for captions as true positives drop.
Errors are estimated on compound metrics with a 5-fold cross-validation, with averages alone shown for the true positive, false positive, and false negative metrics for the sake of clarity.

The distribution of F1 score over time (in increments of 10 years) for figures and figure captions is shown in the bottom panel of \autoref{fig:f1time}.  When compared to the distribution of PDF parsability in figures shown in \autoref{fig:distributionPdfParse}, our method shows less decline toward earlier publication times.

\begin{table*}[!htp]
\begin{center}
\begin{tabular}{|c?cc?cc?cc|}
    \hline
    & \multicolumn{2}{c?}{IOU=0.1} & \multicolumn{2}{c?}{IOU=0.6} & 
    \multicolumn{2}{c|}{IOU=0.8} \\ 
    & figure & caption & figure & caption & figure & caption \\
    \hline
    TP & 96.5\unskip  & 92.6\unskip  & 94.3\unskip  & 88.9\unskip      & 93.3\unskip  & 87.5\unskip   \\   
    FP & 2.9\unskip  & 2.7\unskip  & 5.1\unskip  & 6.4\unskip      & 6.1\unskip  & 7.8\unskip   \\  
    FN & 3.5\unskip  & 6.0\unskip  & 3.5\unskip  & 6.0\unskip      & 3.5\unskip  & 6.0\unskip   \\  
    \thickhline 
    Prec & 97.1$\pm$1.5\unskip  & 97.1$\pm$1.1\unskip  & 94.7$\pm$2.3\unskip  & 93.2$\pm$2.3\unskip      & 93.8$\pm$3.0\unskip  & 91.8$\pm$1.6\unskip   \\  
    Rec & 96.5$\pm$1.6\unskip  & 94.0$\pm$2.7\unskip  & 96.4$\pm$1.7\unskip  & 93.8$\pm$2.8\unskip      & 96.4$\pm$1.0\unskip  & 93.7$\pm$1.5\unskip   \\  
    F1 & 96.7$\pm$0.8\unskip  & 95.5$\pm$1.2\unskip  & 95.5$\pm$1.7\unskip  & 93.4$\pm$1.0\unskip      & 95.1$\pm$1.6\unskip  & 92.7$\pm$1.4\unskip   \\  
    \hline
\end{tabular}
\end{center}
\caption{ Metrics of our main model for different IOU cut-offs with the final test dataset.    There are {\protect490\unskip} figures and {\protect487\unskip} figure captions in the test dataset. AP for figures and figure captions using the COCO  [0.5:0.95:0.05] scheme \citep{coco} is  {\protect91.5\unskip}\% and {\protect87.9\unskip}\%, respectively. TP, FP and FN are shown as percentages of the total instances in each category. Errors are calculated from the standard deviation of a 5-fold cross validation.}
\label{tab:mainResults}
\end{table*}

In what follows, we contextualize these results with comparisons to other methods and datasets, specifically at high degrees of localization.

\subsection{Highly-localized Page Objects - Relevance for Figure extraction} \label{section:high}

Before comparing our models to others, we begin by defining what is meant by ``highly-localized" in the context of document layout analysis.

Prior efforts have highlighted the issues in directly translating object detection (and segmentation) metrics to document layout analysis \citep{wick_fully_2018}, and in some cases have developed new metrics specifically for document layout analysis and it's related processes \citep{pletschacher_page_2010}.
Particular to our YOLO-based application, while the intersection-over-union metric used in object detection effectively weights all of the intersection area equally, eye tracking studies suggest that the key elements of interpreting a graph include the graph x and y axis and their labels \citep[e.g.][]{exampleeyetracking} which tend to be at the edges of figures.
Thus, any translation of object detection metrics, in this case the IOU measurement, to document layout analysis should involve the quantification of how effective the metric is at capturing the number of times these vital figure elements at the edges of the bounding boxes are missed.

\begin{figure*}[!htp]
\centering
\includegraphics[width=0.99\textwidth]{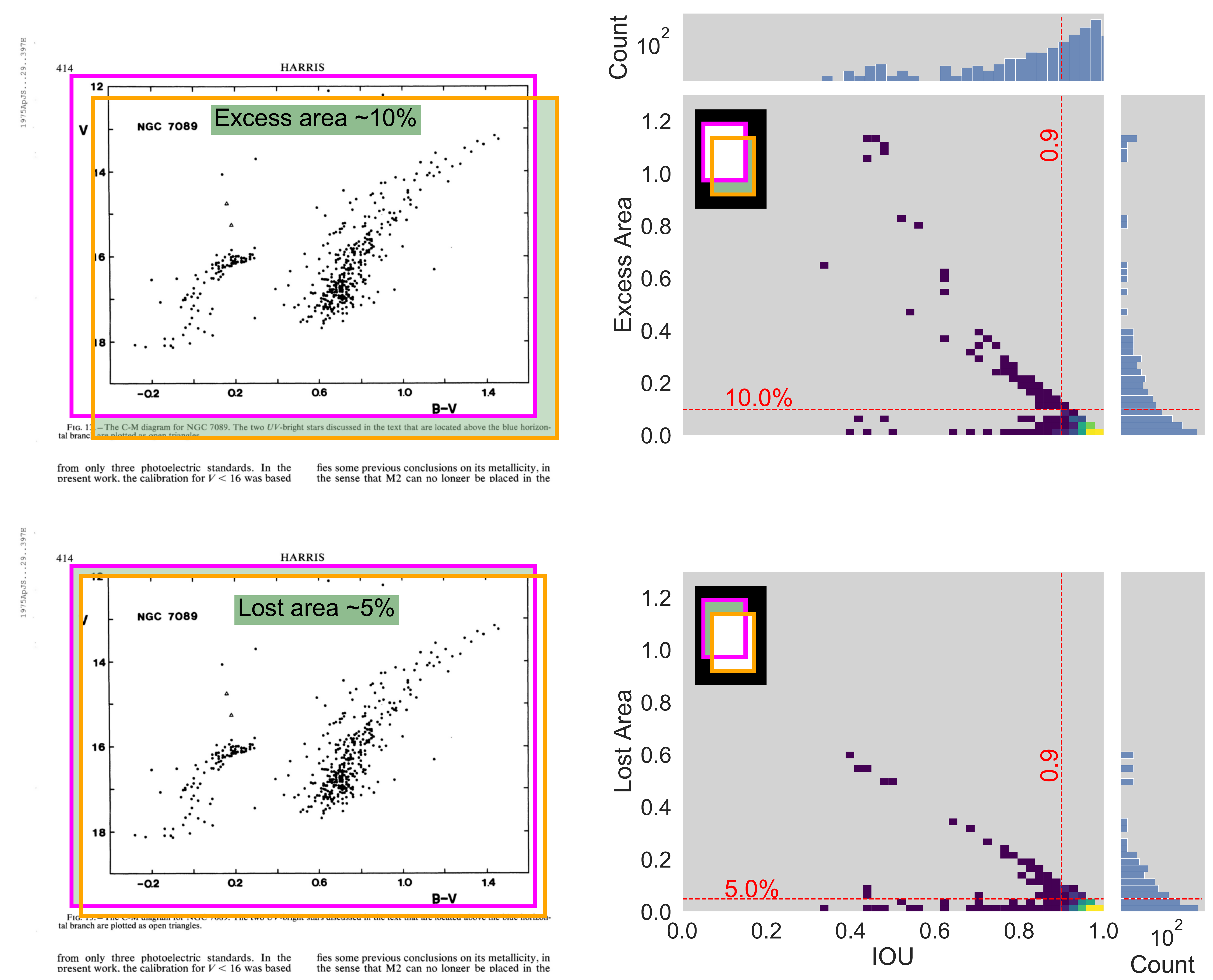}
\caption{\rt{Examples of ``excess area" (upper left panel) and ``lost area" (bottom left panel) along with d}istributions of area in excess (upper \rt{right} panel) and lost (lower \rt{right} panel) \rt{across} \rt{as a percentage of the ground-truth box in} all IOU's of true-found box pairs of figures in the training and validation datasets.  Shaded green regions in the diagrams in the upper left portions of \rt{jointplots also} show the definitions of these areas in relation to true (magenta) and found (orange) boxes.  The IOU chosen for ``highly-localized" page objects is 0.9 (see text for further details).}
\label{fig:ioucutoff}
\end{figure*}

While our current annotation methods do not independently track x and y axis labels, we estimate this effect through the ``area in excess" and ``area lost" from a ground truth figure at a specific IOU with a found box.
\rt{These areas are depicted by an example for one figure in the left panels of \autoref{fig:ioucutoff}.  
The pixels residing outside the ground truth box (magenta boxes) and inside the found box (orange boxes) are summed to calculate the ``area in excess" (green shaded region of diagram in upper left panel of \autoref{fig:ioucutoff}) while the sum of all pixels inside the ground truth box but outside the found box are the ``area lost" (green shaded region of diagram in lower left panel of \autoref{fig:ioucutoff})
For an area-in-excess of $\sim$10\% of the ground-truth box's area, portions of the figure caption are included in the figure box (green shaded area, upper left panel).  While this included information is not an ideal addition to the extracted figure, it is unlikely to cause confusion to a viewer if included on a hosting website like AIE, or in other figure-analysis applications (e.g.\ in datasets used to study color distributions and other elements of  scientific figures \citep{giannakopoulos2015visual,chartanalysis}). In contrast, we show the significant effects of an area-lost of $\sim$5\% by the green shaded region in the lower left panel.  For larger offset between true and found boxes, it is likely the y-axis label would not be included in the extracted figure. Prior research has shown the axis labels are vital to the understanding of a figure and therefore, there exclusion would render the extracted figure unusable to a reader \citep[e.g.][]{exampleeyetracking}.  }

\rt{These examples guide our selection of acceptable cut offs for information in ``excess" and information that is ``lost" from our figures.  
Given our application of hosting on the AIE platform, our focus is on minimal information loss, while information in excess is a secondary concern.
Thus we select 10\% as our acceptable cut-off for the excess area and 5\% for the loss area.}

\rt{We show these cut-offs for the distribution of area-in-excess and area-lost for all of our true-found boxes as a function of the calculated IOU of these pairs in the jointplots of the right panels in \autoref{fig:ioucutoff}.  
These distributions do not include found (true) boxes which do not have a true (found) pair. As shown by the comparison between the excess area (upper right) jointplot and area lost (lower right) jointplot, the distributions over IOU are similarly shaped with the area lost being overall lower than the area in excess.
This is not unexpected, given that many parts of our post processing described in \autoref{section:postprocessing} involve enlarging the boxes around figures.
While \autoref{fig:ioucutoff} only accounts for true-found pairs for figures, the excess and loss area distributions are similar for captions and are omitted here for brevity. }

\rt{In horizontal dashed lines of the right panels of \autoref{fig:ioucutoff}, we show the proposed estimates for cutoffs of acceptable area-in-excess (10\%) and area-lost (5\%).  The majority of the distributions in the excess/lost area lie within these cut-offs, as shown by the lines in the marginal histograms at the top and sides of the jointplots in \autoref{fig:ioucutoff}.}

Once these cut-offs have been chosen, we calculate the minimum IOU which contains 90\% of the data within these cut-offs in order to avoid any outliers at very low IOU which nonetheless have small values of area in excess or lost.
For both area in excess and lost, this results in an IOU of $\sim0.9$ as shown by the vertical lines in both panels and the top marginal IOU plot of \autoref{fig:ioucutoff}.
\rt{We stress that at this stage, these numbers are only estimates and the exact appropriate cut-offs when translating object detection metrics to document layout analysis is the subject of future work.}

In what follows,  we use IOU~$=0.9$ as our definition of an intersection-over-union metric for a ``high degree of localization" an use this cut-off to quantify the robustness of our model in comparison to others.


\subsection{Benchmarks for highly-localized page objects (IOU=0.9)} \label{section:benchandgen}

As the ultimate goal of our method is the extraction of figures and their captions from scanned pages, we quantify how well our model performs on our dataset for a high degree of localization (using our definition \rt{of} an IOU cut-off of 0.9).
We find F1 scores of 90.9\% (92.2\%) for figure (caption) detections at an IOU of 0.9 as shown in the last row and column of \autoref{tab:otherModelsOurData}.  

To facilitate comparison with other document layout analysis methods we compare our method in two ways -- both by quantifying how well other models perform on our dataset and by comparing how well our model performs on other document layout analysis datasets. 

\rt{The chosen data for this comparison are electronic thesis and dissertations (ETDs) from the \textsf{ScanBank} dataset and the article pages from the PubLayNet dataset, which represent two common methods of scientific literature (thesis and refereed publications, respectively), making them frequent targets for digitization efforts in science for many years \citep{nagy_prototype_1992,scanbank,scanbankthesis,pdffigures2,maskrcnnDocbank,binmakhashen_document_2019,jiang2022,silajev2022}.    }

\begin{figure*}[!htp]
\centering
\includegraphics[width=0.9\textwidth]{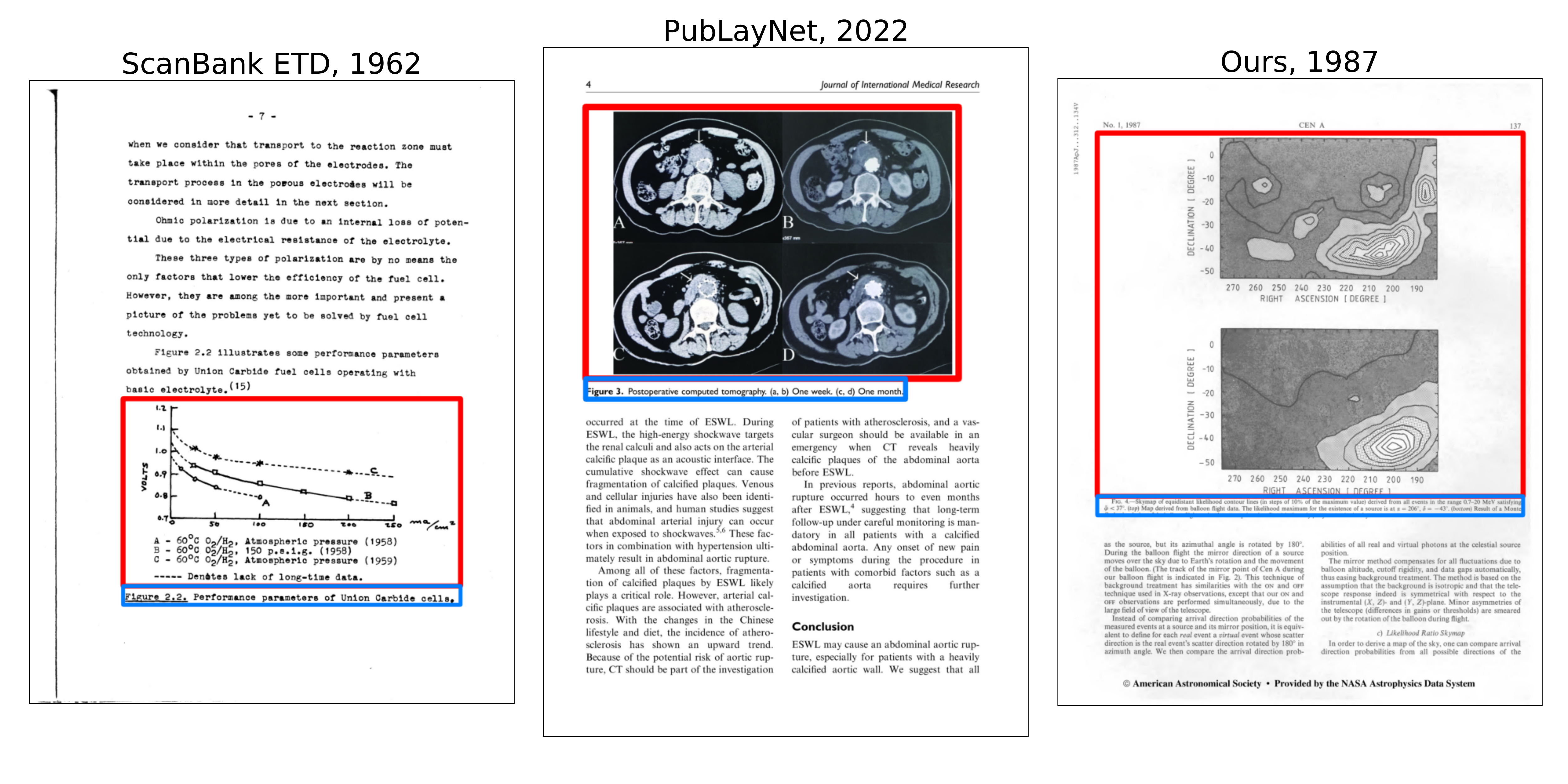}
\caption{\rt{Example pages from the \textsf{ScanBank} collection of historical ETDs \citep{scanbank} (left panel), the PubLayNet dataset \citep{publaynet} (middle panel), and our dataset (right panel).  These panels show the relative differences (font, line spacing, page size) and similarities (figure and caption pairs and placement) between these three datasets.  These pages have been annotated with our definitions for figures (red boxes) and figure captions (blue boxes).}}
\label{fig:scanpub}
\end{figure*}

\rt{ETDs represent the culmination of the graduate work (Masters and/or Ph.D.) of a scientist and their study is an active and growing field within the information sciences \citep{gupta2019}.  The ETDs within the \textsf{ScanBank} dataset (published with the \textsf{ScanBank} object detection model \citep{scanbankthesis}) are collected from the MIT DSpace repository\footnote{\rt{MIT DSpace is a repository which houses ``peer-reviewed articles, technical reports, working papers, theses, and more" \url{https://dspace.mit.edu/}}}, but are limited to those published prior to 1990 to assure they are in raster, not vector format \citep{scanbank}.  The \textsf{ScanBank} dataset contains $\sim$10k pages from 70 ETDs from the years 1900-1990, from a total of 36 different fields including those from STEM (e.g.\ Civil/Mechanical Engineering, Physics, Chemistry) and other fields (e.g.\ Humanities, Political Science, Modern Languages and Linguistics).  An example page containing a figure and caption pair from an ETD in this dataset is shown in the left panel of \autoref{fig:scanpub}.  In comparison with our data (an example page shown in the right panel of \autoref{fig:scanpub}, the ETD text is more widely spaced and the font, line, and page spacing is larger.  While there can be many different forms of ETDs from different fields, the larger line and page spacing is typical for this type of scientific product. }

\rt{In contrast to the visual differences between our dataset and the \textsf{ScanBank} dataset, an example page from the PubLayNet dataset \citep{publaynet} is shown in the middle panel of \autoref{fig:scanpub}.  The PubLayNet dataset is significantly larger than both our and the \textsf{ScanBank} datasets at $\sim$360k pages mined from over one million ``post-digital" articles hosted on PubMed\footnote{\url{https://pubmed.ncbi.nlm.nih.gov/}}.  Beyond the difference in page size, the PubLayNet dataset is visually similar to ours, as shown in the comparison between the middle and right panels of \autoref{fig:scanpub} -- fontsizes, page layout, and line spacing are similar in both panels.}

\rt{Both the \textsf{ScanBank} and PubLayNet datasets include page layout annotations.  Classes of figures, tables and their captions are included in the \textsf{ScanBank} dataset while text, title, list, table and figure are included in the PubLayNet dataset.  The difference in the number of articles between these two datasets highlights an ongoing issue when training models to digitize pre-digital literature -- historical document layout analysis datasets require hand-annotations \citep[\textsf{ScanBank}][]{scanbank}, while newer articles which are post-digital can be mined from their online storage formats \citep[e.g.\ XML, PubLayNet][]{publaynet}. }

\rt{These datasets overlap with different aspects of our dataset.  \textsf{ScanBank}'s age distribution is similar to ours, but differs in its larger diversity of represented fields and publishing venues.  The pages in the PubLayNet dataset are formatted more similarly to our dataset, and belong to a STEM field, as is true of our dataset, however they are from articles published more recently than those within our dataset.  Given these differences and similarities, in what follows, we compare not only how well our model performs on our dataset, but how models trained on these datasets perform on our dataset.}

\autoref{tab:otherModelsOurData} shows how other deep learning models fair on our final test dataset.
Here, we use \textsf{ScanBank} \cite{scanbankthesis,scanbank} (based on \textsf{DeepFigures} \cite{deepfigures} which is trained on the \textsf{ScanBank} ETDs) and a version of \textsf{detectron2} \cite{wu2019detectron2} trained on the PubLayNet dataset \cite{publaynet}.

\textsf{ScanBank} and \textsf{detectron2} are used for comparison as they are applied to raster-formatted articles (as opposed to vector-based methods like \textsf{pdffigures2} \cite{pdffigures2} which, as discussed in \autoref{section:data} results in low accuracies for our dataset. 
Both \textsf{ScanBank} and \textsf{detectron2} do not share our definitions of figures and figure captions exactly, thus to facilitate comparison we make some approximations and assumptions.

As discussed in \cite{scanbankthesis}, \textsf{ScanBank}'s figures are defined as encompassing the figure caption, while our figure definitions exclude the caption.  In order to compare with our results, we initially performed metric calculations by re-defining our true figure boxes as the combination of figure and figure caption boxes when figure captions are associated with a figure.  However, we found that if we instead use our definitions of figures and captions, metrics from detections made with \textsf{ScanBank} are optimized, thus we use our definitions of figures and captions for all comparisons with this model. 
As \textsf{detectron2} does not find caption boxes specifically, but rather localizes generic ``text" boxes, we define \textsf{detectron2}-detected figure captions as those boxes with centers which are closest to a found figure's center.
To test the effects of our post-processing methods alone, we apply a subset of our post-processing steps to the results generated from both \textsf{ScanBank} and \textsf{detectron2} which we show for comparison to our method with and without post-processing in \autoref{tab:otherModelsOurData}. When applying post-processing to these other models' results, we use only up to the ``Step 5" described in \autoref{section:postprocessing} as we found this optimized the metrics for \textsf{ScanBank} and \textsf{detectron2} reported in \autoref{tab:otherModelsOurData}.

As shown in \autoref{tab:otherModelsOurData}, \textsf{ScanBank} does not perform well on our final test dataset with or without post-processing.
In particular, \textsf{ScanBank} does not detect captions reliably as the false negative rate (FN) is high.  
Additionally, there is a large portion of both figures and figure captions which are either erroneously detected, or not well localized as shown by the high false positive (FP) rate.
This is somewhat expected as the ETD format is visually distinct from the articles in our dataset, including different fonts and caption placements.
When post-processing is applied the metrics for figure captions improve significantly with an increase of $\approx$30\% in true positive (TP) rate and decrease of $\approx$20\% in FP.

For the \textsf{detectron2} model without post-processing, TP rates are slightly higher than \textsf{ScanBank}'s,  
however FP rates remain comparable to \textsf{ScanBank}'s. 
FN rates are lower than our model (both with and without post-processing) by a few percentage points, likely due in part to the known differences in error profiles between YOLO-based (ours) and Mask-RCNN-based (\textsf{detectron2}) object detection models \cite{yolov1}.
Post-processing makes a large improvement on the TP rate for captions, increasing it by $\approx$35\% and decreasing FP by $\approx$40\%.  There is a modest increase in TP of $\approx$10\% for figures as well when post-processing is applied.

Post-processing (using all steps) has the largest effect on our model's results -- increasing TP rates of figures and captions by $\approx$25\% and $\approx$60\%, respectively.
This is not surprising as our post-processing method was developed using our scanned page training data.  Additionally, we employ a YOLO-based model which is used for detecting bounding boxes, not a segmentation method that would tend to produce larger TP rates at higher IOU thresholds -- the post-processing pipeline ``mimics" segmentation by changing box sizes \rt{to fit more closely the} precise locations of caption words and figures, increasing overlap IOU. 

Taken together, the results of \autoref{tab:otherModelsOurData} suggest that other models generalize to our dataset at best moderately at high IOU, and only with application of our post-processing pipeline.
Because our post-processing steps require not only grayscale scanned pages, but their OCR outputs, this additional overhead (of both producing the OCR and post-processing steps) greatly reduces the gains in page processing speeds achieved with these other methods.

\begin{table*}[!htp]
\begin{center}
\begin{tabular}{|c?cc?cc?cc?cc?cc?cc|}
    \hline
    & \multicolumn{2}{c?}{ScanBank}  & \multicolumn{2}{c?}{ScanBank}
    & \multicolumn{2}{c?}{detectron2$^\star$} & \multicolumn{2}{c?}{detectron2$^\star$} 
    & \multicolumn{2}{c?}{Ours} & \multicolumn{2}{c|}{Ours} \\
    & \multicolumn{2}{c?}{No PP} 
    & \multicolumn{2}{c?}{w/PP} 
    & \multicolumn{2}{c?}{No PP} & \multicolumn{2}{c?}{w/PP} 
    & \multicolumn{2}{c?}{No PP} & \multicolumn{2}{c|}{w/PP} \\
    & fig
    & cap & fig & cap & fig & cap$^\dagger$ & fig & cap$^\dagger$ & fig & cap & fig & cap \\
    
    \hline
    TP &  
    69.9\unskip & 29.0\unskip &
    69.3\unskip & 52.8\unskip &
    72.0\unskip & 46.4\unskip &
    81.0\unskip & 80.9\unskip &
    58.2\unskip & 23.2\unskip &
    85.7\unskip & 86.7\unskip \\
    
    FP & 71.4\unskip & 28.8\unskip &
    43.6\unskip & 8.7\unskip &
    41.8\unskip & 68.2\unskip &
    27.1\unskip & 22.4\unskip &
    45.3\unskip & 82.3\unskip &
    13.7\unskip & 8.6\unskip \\

    FN &  1.7\unskip & 42.8\unskip &
    2.5\unskip & 40.7\unskip &
    0.6\unskip & 1.6\unskip &
    1.2\unskip & 4.9\unskip &
    3.1\unskip & 5.1\unskip &
    3.5\unskip & 6.0\unskip \\

    \thickhline 
    Prec &  49.5\unskip & 50.2\unskip &
    61.4\unskip & 85.9\unskip &
    63.3\unskip & 40.5\unskip &
    74.9\unskip & 78.3\unskip &
    56.2\unskip & 22.0\unskip &
    86.2\unskip & 90.9\unskip \\

    Rec &  97.6\unskip & 40.4\unskip &
    96.5\unskip & 56.5\unskip &
    99.2\unskip & 96.6\unskip &
    98.5\unskip & 94.3\unskip &
    95.0\unskip & 81.9\unskip &
    96.1\unskip & 93.6\unskip \\
    
    F1 &  65.7\unskip & 44.8\unskip &
    75.0\unskip & 68.1\unskip &
    77.2\unskip & 57.1\unskip &
    85.1\unskip & 85.6\unskip &
    70.6\unskip & 34.7\unskip &
    90.9\unskip & 92.2\unskip \\

    \hline
\end{tabular}
\end{center}
\footnotesize{$^\star$ The tested version of \textsf{detectron2} is trained on the PubLayNet dataset \cite{wu2019detectron2}.}\\
\footnotesize{$^\dagger$ Here, captions are the ``text" classified box closest to the center of a figure.}\\
\caption{Performance metrics for 
\textsf{ScanBank} \cite{scanbankthesis,scanbank} 
and \textsf{detectron2} \cite{wu2019detectron2} for our final test dataset. IOU is 0.9. TP, FP, FN are in percentages of total true instances. Models with post-processing (``w/PP") and those without (``No PP") are shown for comparison. No retraining or transfer learning of \textsf{ScanBank} or \textsf{detectron2} have been done with our dataset. Errors from a 5-fold cross validation on all metrics are $\sim$1-2\%.}
\label{tab:otherModelsOurData}
\end{table*}


This lack of generalizability is a known problem in the field of document layout analysis \citep[e.g.][]{surveydeeplearning} and our model is no exception.
\autoref{tab:ourModelOtherData} quantifies how well our model performs on a collection of ETDs from the \textsf{ScanBank} ``gold standard" dataset \citep{scanbankthesis,scanbank} and a selection of PubLayNet's non-commerical article pages (non-commercial in order to access the high resolution scans and perform the OCR needed for our method) \citep{publaynet}.
Here, we show how well our model performs on this set of hand-annotated ETD and PubLayNet pages using our definitions of figure and figure caption\rt{s, as shown by the red (figure) and blue (caption) boxes in the left (ScanBank) and middle panels (PubLayNet) of \autoref{fig:scanpub}}.
Additionally, for comparison, we show the results of \textsf{detectron2} (\textsf{ScanBank}) on the same set of pages from the PubLayNet (ETDs) dataset in parenthesis in \autoref{tab:ourModelOtherData}, using the same approximations for figure caption boxes used in \autoref{tab:otherModelsOurData}.

\begin{table*}[!htp]
\begin{center}
\begin{tabular}{|c?cc?cc|}
    \hline
    & \multicolumn{2}{c?}{PubLayNet (Non.Comm.)}  & \multicolumn{2}{c|}{ETDs (ScanBank)}  \\
    & figure & caption$^\dagger$ & figure & caption \\
    & 207\unskip & 201\unskip & 197\unskip & 140\unskip \\
    \hline
    TP & 55.1\unskip (83.1\unskip) & 50.2\unskip (55.2\unskip) 
    & 20.8\unskip (32.5\unskip) & 9.3\unskip (0.7\unskip) \\

    FP & 45.4\unskip (25.6\unskip) & 23.4\unskip (56.7\unskip) & 89.1\unskip (84.8\unskip) & 12.4\unskip (10.7\unskip) \\

    FN & 12.1\unskip (1.9\unskip) & 29.4\unskip (1.5\unskip) & 32.1\unskip (19.3\unskip) & 83.4\unskip (90.0\unskip) \\
    \thickhline
    
    Prec & 54.8\unskip (76.4\unskip) & 68.2\unskip (49.3\unskip) & 18.9\unskip (27.7\unskip) & 42.9\unskip (6.2\unskip) \\
    Rec & 82.0\unskip (97.7\unskip) & 63.1\unskip (97.4\unskip) & 39.3\unskip (62.7\unskip) & 10.1\unskip (0.8\unskip) \\
    F1 & 65.7\unskip (85.8\unskip) & 65.6\unskip (65.5\unskip) & 25.5\unskip (38.4\unskip) & 16.3\unskip (1.4\unskip) \\
   
    \hline
\end{tabular}
\end{center}
\footnotesize{$^\dagger$ For \textsf{detectron2} we assume a box classified as ``text`` which is closest to the center of a found figure is it's associated caption.}\\
\caption{Performance metrics for our model's performance on the non-commercial pages of the PubLayNet dataset \citep{publaynet} and pages from ETDs in the ``gold standard" dataset of \textsf{ScanBank} \citep{scanbank,scanbankthesis} for IOU=0.9.  For comparison included in parenthesis are metrics for models trained on these datasets -- \textsf{detectron2} for the PubLayNet dataset and \textsf{ScanBank} for the ETD dataset.  Comparisons should be taken as first estimates -- see text for further details.}
\label{tab:ourModelOtherData}
\end{table*}

Both \textsf{ScanBank} and \textsf{detectron2} perform better on \rt{the} figures \rt{from their respective training datasets} than our model, with increases in true positive rates of $\approx$10\% and $\approx$30\% over our detections, respectively. 
False positive rates tend to be comparable (\textsf{ScanBank}) or higher (by $\approx$20\% for \textsf{detectron2}) and false negative rates are higher as well.

Results for figure captions tend to be comparable (\textsf{detectron2}) or better (\textsf{ScanBank}) using our model, however we again caution here that the definitions of captions and figures is not constant across all datasets making these comparisons only estimations.

While these results suggest that our model may be more generalizable for figure captions, \rt{further tests on a more datasets would strengthen this conclusion.} 
\rt{However, the authors caution that the ability of {\textit{any}} present-day document layout analysis model to generalize well is typically limited \citep{podreview1,naiman2023}.  This lack of generalizability does not appear to depend on training dataset size \citep{naiman2023}.  Compounding the problem, the definitions of figures, captions and other page objects can differ between fields and indeed the annotators of documents within the same field \citep{younas2019,doclaynet}, making even ``brute force" hand-classification of diverse sets of documents with high accuracy a non-trivial problem \citep{doclaynet}. }

\section{Discussion and Future Work} \label{section:future}
This paper has focused on the localization of figures and figure captions in astronomical scientific literature.
We present results of a YOLO-based deep learning model trained on features extracted from the scanned page, \textsf{hOCR} outputs of the \textsf{Tesseract} OCR engine \cite{tesseract}, and the text processing outputs from spaCy \cite{spacy2}.  

As our dataset comprises both vector-PDF and raster-PDF formats, we test the ``PDF parsability" of the articles by passing them through the PDF-mining software \textsf{pdffigures2} \citep{pdffigures2} and \textsf{GROBID} \citep{GROBID}.  We find a {\textit{maximum}} of \rt{$\approx 33\%$} of our articles are parsable, thus motivating our approach of an object-detection based method for finding figures and captions.

We spend considerable effort to precisely define the classes of ``figure" and ``figure caption" to both avoid the differences of classifications that can be present in other datasets \cite[e.g.][]{younas2019} and to align with our goals of not only localizing these objects but extracting them from scanned pages to be hosted separately from their articles of origin.

Through ablation experiments we find the combination of the page and \textsf{hOCR} properties of (grayscale,  ascenders, descenders, word confidences, fraction of numbers in a word, fraction of letters in a word, punctuation, word rotation and spaCy POS) maximize our model's performance at detecting figures and their captions.
When compared to other deep learning models popular for document layout analysis (\textsf{ScanBank} \cite{scanbank,scanbankthesis} and \textsf{detectron2} \cite{wu2019detectron2}) we find our model performs better on our dataset, particularly at the high IOU thresholds (IOU=0.9) and especially for figure captions.

This IOU cut off is motivated by an analysis of the ``area lost" and ``area in excess" for true-found box pairs and we thus define the IOU=0.9 cut-off as the definition of ``highly-localized" page objects in the application of the YOLO-based object detection model to our document layout analysis problem.
In line with our extraction goals, our model has relatively low false positive rates, minimizing the extraction of erroneous page objects.  

Similar to the low generalization of other deep learning models to our dataset at high IOU, our model does not generalize well for the detection of figures in a subset of \textsf{ScanBank}'s collection of ETDs \citep{scanbank} and PubLayNet's Non-Commercial scanned pages \citep{publaynet} used to train the version of \textsf{detectron2} used in our comparisons.  Our model generalizes significantly better for the detection of figure captions, showing comparable or higher true positive rates and lower false positive rates.  Additionally, we show that our post processing pipeline increases the performance of all models, especially for the detection of figure captions. We caution however these comparisons are estimates given the different definitions of figures and captions in our model in comparison to others.

\rt{Taken together, our work in first carefully defining the classes of page objects, then defining ``high-localization" and testing the generalizability of our models along with those of \textsf{ScanBank} and \textsf{detectron2} highlight the need within the document layout analysis community to consider carefully about how we apply the methods of object detection and segmentation to the extraction of page objects.  It is vital that we first define what {\it information} we intend to extract before we quantify how accurately we have performed the extraction.  This is in contrast with other works which typically use mAP or an IOU=0.8 as their metric of comparison without quantifying how this translates into the information lost from the extracted page object \citep{younas2019,yolo1,yang2017layout,maskrcnnDocbank,maskrcnncdec,fcnncharts}.}

Our work relies on a relatively small set of scanned pages ($\sim$6000).  While the results here for figures and captions surpass the estimates of $\sim$2000 instances per class required for training YOLO-based models \cite{bochkovskiy2020yolov4,Wang_2021_CVPR} our data contains many edge cases of complex layouts and we expect more data to improve results for these pages.  \rt{In addition, our definitions do not link together figures that are spread in panels across multiple pages (e.g.\ Fig 1a and Fig 1b on separate pages).  Linking such pages requires the extraction of captions and the denoising of their OCR results with post-correction methods \citep[e.g][]{ramirez-orta_post-ocr_2022} and is the subject of ongoing work \citep{morgan}.}  

As our model relies on more than three feature channels, transfer learning on pre-trained YOLO-based models is less straight forward, but nonetheless could be a way to make use of our small dataset in future work.

Additionally, our current methodology does \rt{not test the efficacy} of popular image processing features (e.g. connected components \cite{younas2019}) or loss functions/processing techniques that are ``non-standard" for YOLO-based methods \cite{non_nms} \rt{on our dataset}.  We also use the ``standard" spaCy package for linguistic feature generation, instead of a science-specific version of spaCy (e.g. ScispaCy\footnote{\url{https://allenai.github.io/scispacy/}.}) Future testing with the inclusion of these features may increase our model performance.

While all of our models converge within 150 training epochs, this is without the inclusion of any data augmentation.  
As our model uses not only grayscale but \textsf{hOCR} properties, typical data augmentation procedures (e.g. flipping, changes in saturation) are not appropriate for all feature layers.  However, it is likely that correctly-applied data augmentation \rt{(e.g.\ grayscale-layer contrast modifications)} will increase our model's \rt{accuracy above the metrics reported here}. 
Our work would further benefit from \rt{both} a future large hyperparameter tuning study beyond the several values of learning rate tested in this paper\rt{, and additional feature selection analysis, as permitted within any computational constraints}.

Finally, given the difference in error profiles between the YOLO-based method presented here and other Mask-RCNN/Faster-RCNN based \cite{yolov1} document layout analysis models (e.g. \textsf{detectron2}), it is likely that an ensemble model using both methods would further increase model performance.

\rt{While this work has been completed with an astronomy-specific dataset, it has shown some promise to greater generalizability than other models \citep[e.g.][]{scanbank,publaynet,wu2019detectron2}, however the accuracy is still far below that necessary for a deployable solution at scale for historical literature in other fields. As suggested in prior work \citep{naiman2023}, the ``generalizability problem" will likely be solved by the careful definitions of page object classes for large-scale annotation campaigns, and quantification of ``high localization" in combination with models like the one which was developed in this work. }


This work is supported by a Fiddler Fellowship and a NASA Astrophysics Data Analysis Program Grant (20-ADAP20-0225). 

\bibliographystyle{sn-mathphys}
\bibliography{references}

\section{Appendix} \label{section:appendix}
\begin{figure*}[!htp]
\centering
\includegraphics[width=0.99\textwidth]{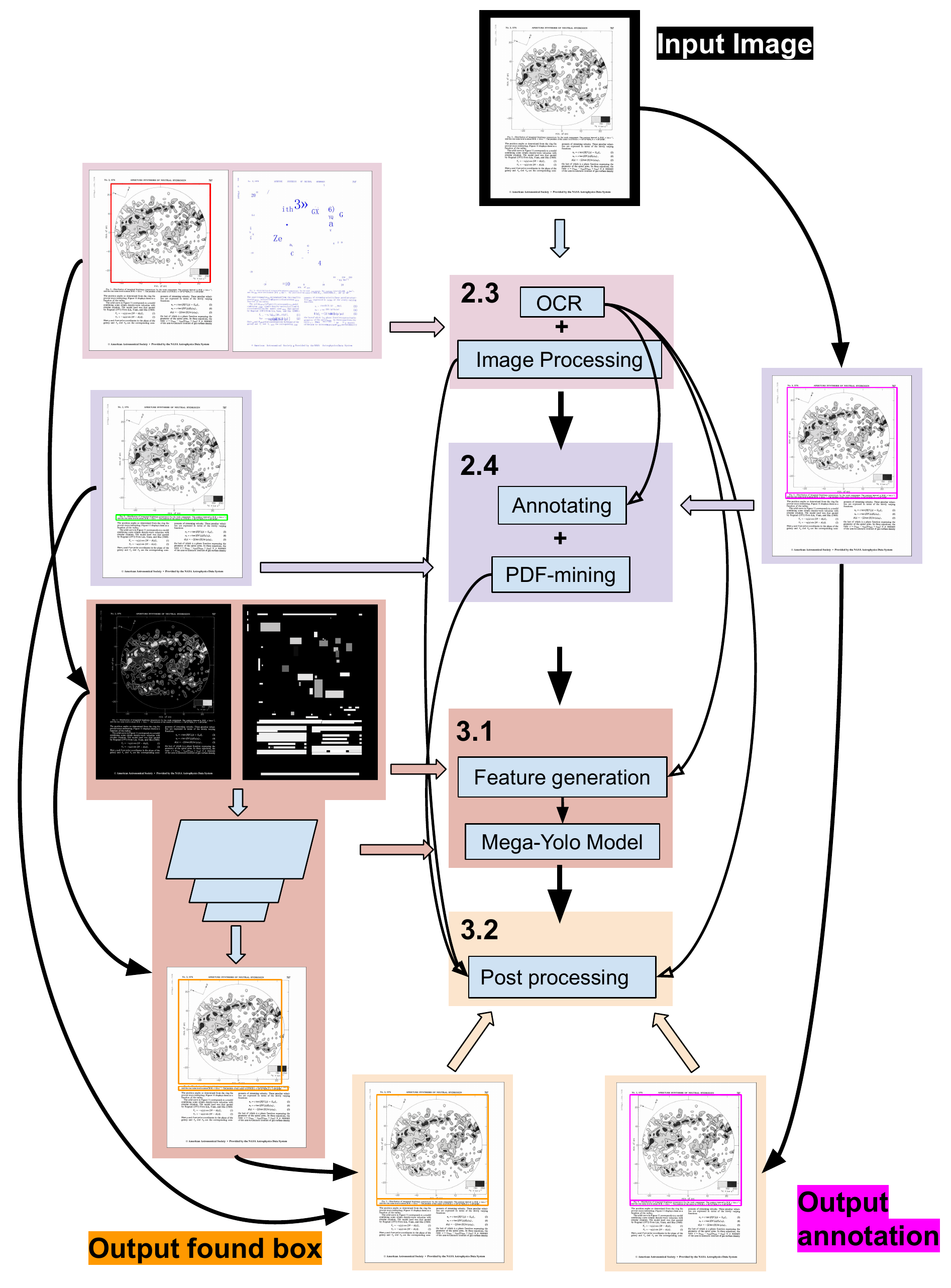}
\caption{\rt{Detailed architecture diagram showing how a scanned page is processed through our full model pipeline (left panels) including OCR'ing and image processing, PDF-mining, feature generation and the subsequent feeding of features into the Mega-Yolo model, with model outputs feeding into the final found boxes. On the right we depict the annotation process performed on the scan page in order to generate ground truth models for comparision with our model's found boxes.}}
\label{fig:pipelineexpand}
\end{figure*}

\rt{Full architecture diagram, expanded from \autoref{fig:pipeline} is shown in \autoref{fig:pipelineexpand}. The scanned page is first processed with the \textsf{Tesseract} OCR engine and image processing is applied to look for any square shapes on the page (left pink panel, \autoref{section:imageprocessing}).  Then PDF-mining is performed to find any figures or captions, though in practice only a small fraction of scans contain PDF-parsable page objects, and these are typically captions only (left purple panel, \autoref{section:annotations}).  Then, features are generated from the grayscale and OCR data, which are then processed through the Mega-Yolo Model which generates the raw found boxes (left red panel, \autoref{section:modeldesign}).  Finally, the PDF-mining-found captions, image-processing-found figure boxes, and OCR word bounding boxes are used to clean the raw found boxes resulting in the output found box (lower left orange panel, \autoref{section:postprocessing}).}

\rt{To generate ground-truth boxes, the hand annotating of the same page is performed (right purple panel). Then the OCR word bounding boxes are used to augment the hand-annotated boxes to generate accurate ground-truth figure and caption boxes (bottom right orange panel).}

\end{document}